# Stabilization and Re-excitation of Sawtooth Oscillations due to Energetic Particles in Tokamaks

H.X. Zhang[1], H.W. Zhang[2,*], Z.W. Ma[1,*], J.X. Huang[1], W. Zhang[1]

[1] Institute for Fusion Theory and Simulation, Zhejiang University, 310027 Hangzhou, China

[2] Max Planck Institute for Plasma Physics, Boltzmannstr. 2, 85748 Garching b. M., Germany

## Abstract

Sawtooth oscillations, driven by internal kink modes (IKMs), are fundamental phenomena in tokamak plasmas. They can be classified into different types, including normal sawteeth, small sawteeth, and in some cases, evolving into the steady-island state, each having a different impact on energy confinement in fusion reactors. This study investigates the interaction between sawtooth oscillations and energetic particles (EPs) using the initial-value MHD-kinetic hybrid code CLT-K, which can perform long-term self-consistent nonlinear simulations. We analyze the redistribution of EPs caused by sawtooth crashes and the effect of EPs on sawtooth behavior and type transitions. The results show that co-passing EPs tend to re-excite sawtooth oscillations, extending their period, while counter-passing EPs promote the system evolution toward small sawteeth, potentially leading to the steady-island state. Additionally, we provide a physical picture of how EPs influence sawtooth type through the mechanism of magnetic flux pumping. We demonstrate that the radial residual flow in the core plays a crucial role in determining the reconnection rate and sawtooth type. Moreover, we observe new phenomena about couplings of various instabilities, such as the excitation of global multi-mode toroidal Alfvén eigenmodes (TAEs) due to EP redistribution following a sawtooth crash and the excitation of the resonant tearing mode (r-TM) when injecting counter-passing EPs. The study also explores the impact of EP energy and the safety factor profile on the development of stochastic magnetic fields and EP transport. These findings emphasize the necessity of multi-mode simulations in capturing the complexity of EP-sawtooth interactions and provide insights for optimizing sawtooth control in future reactors such as ITER.

* Corresponding Authors: haowei.zhang@ipp.mpg.de, zwma@zju.edu.cn.

## 1. Introduction

Sawtooth oscillations, or simply "sawteeth", are periodic plasma relaxations due to magnetic reconnections commonly observed in tokamaks [1]. They are characterized by a slow rise in the central plasma temperature and density (the "sawtooth ramp") followed by a rapid collapse (the "sawtooth crash") [2]. According to the well-known models by Kadomtsev [3] and Porcelli et al. [4], sawtooth oscillations are typically triggered by $m/n = 1/1$ internal kink modes (IKMs), where $m$ and $n$ are the poloidal and toroidal mode numbers, respectively. Sawtooth crash rapidly flattens the plasma temperature distribution, significantly threatening energy confinement. In future high-performance burning plasmas, such as those envisioned in ITER, uncontrolled sawtooth oscillations could lead to severe stability issues, posing a substantial risk to reliable operation [5].

Sawtooth oscillations can be classified into different types, including normal sawteeth [1] and small sawteeth [6]. Normal sawteeth, sometimes referred to as "monster sawteeth" [7] or "giant sawteeth" [8], characterized by long periods and large crash amplitudes, involve complete magnetic reconnection during the crash, leading to dramatic redistribution of core temperature and density, as described by Kadomtsev's model [3]. They can trigger secondary magnetohydrodynamic (MHD) instabilities such as neoclassical tearing modes (NTMs) [9-11] and edge localized modes (ELMs) [12], severely degrade energy confinement or even cause disruptions [13]. By contrast, small sawteeth have shorter periods, smaller amplitudes, and involve incomplete magnetic reconnection, making their impact on core plasma less disruptive [11]. They help avoid impurity and helium ash accumulation [14], making them the preferred operational mode for future reactors like ITER [5,15]. Under certain conditions, sawteeth may be completely suppressed, allowing the core to maintain a steady-state $m/n = 1/1$ magnetic island and avoid sawtooth crash [16-19]. This phenomenon may be related to the impurity "snake" observed in tokamaks [20-22]. Plasma viscosity [18,19], thermal conductivity [2], plasma beta [16], Hall effect [23], bootstrap current [24], configuration [25], etc., may all influence the type of sawteeth. To effectively control sawteeth [15], understanding the mechanisms driving transitions between different sawtooth types is essential. Previous studies have shown that electron cyclotron current drive (ECCD) [26-30], auxiliary heating [31-33] such as neutral beam injection (NBI) [34-36], and resonant magnetic perturbations (RMPs) [37,38] can alter sawtooth periods and amplitudes, thereby influencing their type.

Sawteeth can strongly interact with energetic particles (EPs) [35,39]. Early theoretical and initial-value simulation studies have shown that sawtooth crash can significantly redistribute EPs, transporting them from the core to outside the $q = 1$ rational surface [39-44]. Further research indicates that passing EPs are more easily redistributed, while significant trapped EPs redistribution mainly occurs when EP energy is above a threshold [43-47]. This energy-selective effect [44] suggests that high-energy alpha particles may remain in the core to sustain heating [48], while lower-energy helium ash can be expelled [49]. However, these studies mainly focus on how sawtooth crash affect EP transport, i.e., treating EPs as test particles. Simulation research including the feedback of EPs on sawtooth behavior using initial-value code simulations is still lacking.

On the other hand, to clarify the mechanism by which NBI can control sawtooth oscillations [36,50-53], a series of studies considered the influence of EPs on the linear stability of the IKM, suggesting that stabilizing the IKM could extend the sawtooth period, while destabilizing the IKM could shorten it. Analytical theories and eigenvalue analyses show that trapped EPs have a stabilizing effect on IKMs [54], while passing particles can either stabilize or destabilize IKMs, depending on their deposition location and motion direction [55-63]. Specifically, co-passing EPs injected on-axis (off-axis) and counter-passing EPs injected off-axis (on-axis) can stabilize (destabilize) IKMs. This conclusion qualitatively explain the asymmetry observed in the influence of co-current and counter-current NBI on sawtooth periods in experiments such as JET [52,64,65] and TEXTOR [36]. Further studies based on the MARS-K code indicate that the finite orbit width (FOW) effect of EPs could reduce this asymmetry [62], but anisotropic thermal transport could enhance it [63]. Additionally, a negative triangular geometry slightly favors IKM destabilization [62]. However, sawteeth are inherently nonlinear phenomena [35], but for influences of EPs on sawtooth behavior, long-term nonlinear self-consistent simulations are still lacking, and there is little analysis from the perspective of sawtooth type transitions.

This study uses the CLT-K code, based on the MHD-kinetic hybrid model, to perform long-term self-consistent nonlinear simulations of the interaction between passing EPs and sawtooth oscillations. The goal is to develop a physical understanding of how EPs affect the nonlinear behavior and type transitions of sawtooth oscillations, with a focus on the relationship between sawtooth period and core radial residual flow (which is crucial for maintaining steady-states [16-18,30,66]) and magnetic reconnection rate at the X-point. We find that injecting co-passing EPs extends the sawtooth period, lowers the residual flow level, and reduces the reconnection rate at the X-point, favoring large-amplitude sawtooth oscillations. Conversely, injecting counter-passing EPs has the opposite effect, promoting the formation of the steady-state island. Analyzing from the perspective of sawtooth type transitions provides insights for optimizing confinement. Additionally, multi-mode simulations in this study reveal new phenomena in the EP-sawtooth interaction, such as multi-$n$ global toroidal Alfvén eigenmode (TAE) excitation due to sawtooth crashes, the appearance of resonant tearing modes, and the appearance and impact of stochastic fields. These phenomena not only significantly affect plasma confinement but also demonstrate the necessity of multi-mode simulations in initial-value codes.

The structure of this paper is as follows: Section 2 introduces the MHD-kinetic hybrid model used in the CLT-K code. Section 3 presents the initial equilibrium and parameter settings. Section 4 presents the results of pure MHD simulations and discusses the characteristics of various types of sawtooth oscillations. Section 5 shows the simulation results on the influence of injected EPs on sawtooth types and analyzes the mechanisms to provide a physical understanding. Section 6 conducts further studies on different parameters and the new phenomena observed in the EP-sawtooth interaction. Finally, Section 7 summarizes and discusses the findings of the work.

## 2. Simulation Model

The CLT-K code [67-69] is an initial value code based on the MHD-kinetic hybrid model [70], used for various EP physics studies [67-69,71,72], and has been ported to run on graphical processing units (GPUs) [73]. It solves the MHD equations in a 3D cylindrical coordinate system $\{R, \phi, Z\}$ (where $R$ representing the major radius, $\phi$ as the toroidal angle, and $Z$ along the vertical direction) to evolve the background plasma, while the contribution of EPs is coupled to the momentum equation through EP pressure tensor $\mathbf{P}_h$ or EP current $\mathbf{J}_h$ (where the subscript 'h' denotes EPs).

The complete set of equations is as follows:

$$\frac{\partial \rho}{\partial t} = -\nabla \cdot (\rho \mathbf{V}) + \nabla \cdot (D \nabla \rho) \tag{1}$$

$$\frac{\partial p}{\partial t} = -\mathbf{V} \cdot \nabla p - \Gamma p \nabla \cdot \mathbf{V} + \nabla \cdot (\kappa_\parallel \nabla_\parallel p) + \nabla \cdot [\kappa_\perp \nabla (p - p_0)] \tag{2}$$

$$\frac{\partial \mathbf{V}}{\partial t} = -\mathbf{V} \cdot \nabla \mathbf{V} + \frac{\mathbf{J} \times \mathbf{B} - \nabla p - (\nabla \cdot \mathbf{P}_h)_\perp}{\rho} + \nabla \cdot (\nu \nabla \mathbf{V})$$

$$= -\mathbf{V} \cdot \nabla \mathbf{V} + \frac{(\mathbf{J} - \mathbf{J}_h) \times \mathbf{B} - \nabla p}{\rho} + \nabla \cdot (\nu \nabla \mathbf{V}) \tag{3}$$

$$\frac{\partial \mathbf{B}}{\partial t} = -\nabla \times \mathbf{E} \tag{4}$$

$$\mathbf{E} = -\mathbf{V} \times \mathbf{B} + \eta(\mathbf{J} - \mathbf{J}_0) \tag{5}$$

$$\mathbf{J} = \nabla \times \mathbf{B} \tag{6}$$

where $\rho$, $p$, $\mathbf{V}$, $\mathbf{B}$, $\mathbf{E}$, and $\mathbf{J}$ are plasma density, plasma pressure, fluid velocity, magnetic field, electric field, and total current density respectively. $\Gamma = 5/3$ is the ratio of specific heat of the plasma, and $D$, $\kappa_\parallel$, $\kappa_\perp$, $\nu$, and $\eta$ are the dissipation coefficients (the diffusivity, parallel and perpendicular thermal conductivity, viscosity, and resistivity). All evolving physical quantities can be decomposed into the initial equilibrium and perturbation components, e.g., $\mathbf{B} = \mathbf{B}_0 + \delta \mathbf{B}$, where the subscript '0' represents the equilibrium and the prefix 'δ' signifies the perturbation. The terms involving '$-p_0$' and '$-\mathbf{J}_0$' in equations (2) and (5) correspond to the heat source and current source for the plasma. These sources allow the pressure and current distributions to return to equilibrium after each sawtooth crash, making the onset of the next sawtooth cycle possible, thereby leading to periodic oscillations in the system. Its MHD version has been benchmarked for consistency with the M3D-C1 code [74], while the MHD-kinetic version has undergone preliminary benchmarking with M3D-K and M3D-C1-K codes [69].

All physical quantities are dimensionless: $x/a \to x$, $t/\tau_A \to t$, $B/B_{m0} \to B$, $\rho/\rho_{m0} \to \rho$, $p/(B_{m0}^2/\mu_0) \to p$, $P_h/(B_{m0}^2/\mu_0) \to P_h$, $V/v_A \to V$, $J/(B_{m0}^2/\mu_0 a) \to J$, $J_h/(B_{m0}^2/\mu_0 a) \to J_h$, $E/(v_A B_{m0}) \to E$, $D/(v_A a) \to D$, $\kappa_\parallel/(v_A a) \to \kappa_\parallel$, $\kappa_\perp/(v_A a) \to \kappa_\perp$, $\nu/(v_A a) \to \nu$, and $\eta/(\mu_0 v_A a) \to \eta$, where $B_{m0}$ and $\rho_{m0}$ are the equilibrium magnetic field and plasma density at the magnetic axis respectively, $a$ is the tokamak's minor radius, $v_A = B_{m0}/\sqrt{\mu_0 \rho_{m0}}$ is the Alfvén speed at the magnetic axis, $\tau_A = a/v_A$ is the Alfvén time, and $\omega_A = \tau_A^{-1}$ is the Alfvén frequency.

To obtain the contribution of EPs, i.e., the EP pressure $P_h$ or EP current $J_h$, we use the Particle-in-cell (PIC) method, representing EPs with a number of markers and evolving their distribution function $f$ in the phase space $\{X, v\}$, where $X$ is the coordinate and $v$ is the velocity of the gyrocenter. In CLT-K, this process is divided into three steps:

**Step 1: Pushing EPs in the electromagnetic field.**

Since we adopt the guiding-center approximation and neglect the finite Larmor radius (FLR) effect of EPs, the particle's magnetic moment $\mu$ is an invariant. Therefore, the markers can be advanced in the simplified four-dimensional phase space $\{X, v_\parallel\}$. The equation is as follows [75]:

$$\frac{dX}{dt} = \frac{1}{B_\parallel^*}(v_\parallel B^* + E^* \times b) \tag{7}$$

$$\frac{dv_\parallel}{dt} = \frac{1}{B_\parallel^*}\frac{q_h}{m_h}B^* \cdot E^* \tag{8}$$

where $q_h$ and $m_h$ are the charge and mass of the EP, $b$ is the unit vector along the direction of the magnetic field (all "perpendicular" and "parallel" are defined with respect to $b$). Here, $B^* = \nabla \times A^*$, $E^* = -\nabla \varphi^* - \partial A^*/\partial t$, where $\varphi^*$ and $A^*$ are called effective electric potential and effective magnetic vector potential. Their relationship with the real electric potential $\varphi$ and magnetic vector potential $A$ is: $\varphi^* = \varphi + \mu B/q_h$, $A^* = A + (m_h/q_h)v_\parallel b$.

**Step 2: Evolving the distribution function of EPs with $\delta f$ method.**

In the CLT-K code, we express the distribution function of EP as a function of three conservation quantities $\{P_\phi, \varepsilon, \Lambda\}$, where $P_\phi = m_h v_\parallel R B_\phi/B - q_h \psi$ is the poloidal canonical angular momentum (where $\psi$ is the poloidal magnetic flux), $\varepsilon = \mu B + m_h v_\parallel^2/2$ is the energy, and $\Lambda = \mu B_{m0}/\varepsilon$ is the pitch angle. Using the $\delta f$ method, only the perturbation of the distribution function is considered. To improve computational accuracy, we apply non-uniform Monte Carlo sampling [76,77] with a weighting factor $g$. Thus, the actual evolution is for the perturbation distribution function $w = \delta f/g$ contributed by each marker. The equation is:

$$\frac{dw}{dt} = -\frac{1}{g}\frac{df_0}{dt} = -\frac{1}{g}\left(\frac{\partial f_0}{\partial P_\phi}\frac{dP_\phi}{dt} + \frac{\partial f_0}{\partial \varepsilon}\frac{d\varepsilon}{dt} + \frac{\partial f_0}{\partial \Lambda}\frac{d\Lambda}{dt}\right) \tag{9}$$

**Step 3: Integrating the distribution function of EPs to obtain the EP pressure or current.**

In the pressure-coupling scheme, we adopt the EP pressure $\mathbf{P}_h$ in Chew-Goldberger-Low (CGL) form [78], that is:

$$\mathbf{P}_h = P_{h\perp}\mathbf{I} + (P_{h\parallel} - P_{h\perp})\boldsymbol{bb} \tag{10}$$

where $\mathbf{I}$ is the unit tensor.

In the current-coupling scheme, the EP current is divided into two components: the guiding-center current (including the current contributions from the EPs' parallel motion, magnetic gradient drift, and magnetic curvature drift, while the contribution from $\boldsymbol{E} \times \boldsymbol{B}$ drift is self-consistently canceled in the momentum equations) and the magnetization current, given by [79]:

$$\begin{aligned} \boldsymbol{J}_h &= \int q_h\left(v_\parallel \boldsymbol{b} + \frac{m_h v_\parallel^2}{q_h B}\boldsymbol{\nabla}\times\boldsymbol{b} + \frac{\mu}{q_h B}\boldsymbol{b}\times\boldsymbol{\nabla}B\right)f\mathrm{d}\boldsymbol{v} + \boldsymbol{\nabla}\times\left(-\int\mu\boldsymbol{b}f\mathrm{d}\boldsymbol{v}\right) \\ &= q_h\widetilde{NV}_\parallel\boldsymbol{b} + \frac{1}{B}(P_{h\parallel} - P_{h\perp})\boldsymbol{\nabla}\times\boldsymbol{b} + \frac{1}{B}\boldsymbol{b}\times\boldsymbol{\nabla}P_{h\perp} \end{aligned} \tag{11}$$

In the above equations (10) and (11),

$$P_{h\parallel} = \int m_h v_\parallel^2 f\mathrm{d}\boldsymbol{v} \tag{12}$$

$$P_{h\perp} = \int \mu B f\mathrm{d}\boldsymbol{v} \tag{13}$$

$$\widetilde{NV}_\parallel = \int v_\parallel f\mathrm{d}\boldsymbol{v} \tag{14}$$

In previous studies [68,69], the strict equivalence of the pressure-coupling and current-coupling schemes has been demonstrated, allowing the EP contribution to be expressed in a unified form:

$$\begin{aligned} \boldsymbol{J}_h \times \boldsymbol{B} &= (\boldsymbol{\nabla}\cdot\mathbf{P}_h)_\perp \\ &= (\boldsymbol{\nabla}P_{h\perp})_\perp + (P_{h\parallel} - P_{h\perp})(\boldsymbol{b}\cdot\boldsymbol{\nabla})\boldsymbol{b} \end{aligned} \tag{15}$$

This explains the second equality in the momentum equation (3).

In CLT-K, the actual evolution is the perturbation of $(\boldsymbol{\nabla}\cdot\mathbf{P}_h)_\perp$ or $\boldsymbol{J}_h \times \boldsymbol{B}$, which can also be divided into two parts: the "$\delta f$ contribution" from EP distribution function evolution, and the "$\delta\boldsymbol{B}$ contribution" from magnetic field perturbations. More details on these contributions and other aspects of the model can be found in Reference [69], and we will also discuss them further in Sections 5.1 and 5.4 of this paper.

## 3. Initial Equilibrium and Parameter Settings

We use NOVA's QSLOVER [80] to generate a circular cross-section tokamak equilibrium with an aspect ratio of $R_0/a = 4$ and a core beta of $\beta_{eq0} = 1.77\%$ as the basis for our study. The safety factor at the magnetic axis is $q_0 = 0.7$ while it is $q_1 = 3.6$ at the boundary. There exists a $q = 1$ rational surface at $r = 0.38$, where the radial coordinate $r$ is estimated as $r \approx \sqrt{\bar{\psi}}$, with $\bar{\psi}$ being the normalized poloidal flux ($\bar{\psi} = 0$ at the core and $\bar{\psi} = 1$ at the boundary). To focus on studying the $m/n = 1/1$ IKM, we design the equilibrium to avoid other MHD instabilities, such as $m \geq 2$ tearing modes. We use the tearing mode instability index $\Delta'$ on rational surfaces as an approximate criterion [81]. For example, in this equilibrium, the instability index for the $m/n = 2/1$ classical tearing mode is $\Delta'_{2/1} \approx -5.05 < 0$, indicating that it is MHD-stable.

For EPs, we use a slowing-down distribution function, given by:

$$f_0 \propto \frac{\mathcal{H}(v_0 - v)}{v^3 + v_c^3} \exp\left(-\frac{\overline{\langle \psi \rangle}}{0.15}\right) \exp\left[-\left(\frac{\Lambda - \Lambda_0}{\Delta \Lambda}\right)^2\right] \tag{16}$$

where $\mathcal{H}(\cdots)$ denotes the Heaviside function, $v_0 = 0.7v_A$ is the birth speed of the EPs, $v_c = 0.5v_0$ is the critical speed, the Larmor radius of EP ($v_\perp = v_0$) is $\varrho_h = 0.07a$, and $\langle \psi \rangle = -P_\phi/q_h + (m_h/q_h)\langle v_\parallel R B_\phi/B \rangle$ denotes the average magnetic flux along the particle orbit (normalized as $\overline{\langle \psi \rangle}$). This study mainly focuses on passing EPs, mimicking tangential NBI injection, with $\Lambda_0 = 0$ and $\Delta \Lambda = 0.3$ chosen. We define the EP beta as $\beta_h = (\beta_{h\parallel} + 2\beta_{h\perp})/3$ [82] and denote the initial core EP beta as $\beta_{h0}$, which is treated as a tunable parameter. The equilibrium profile is shown in Figure 1.

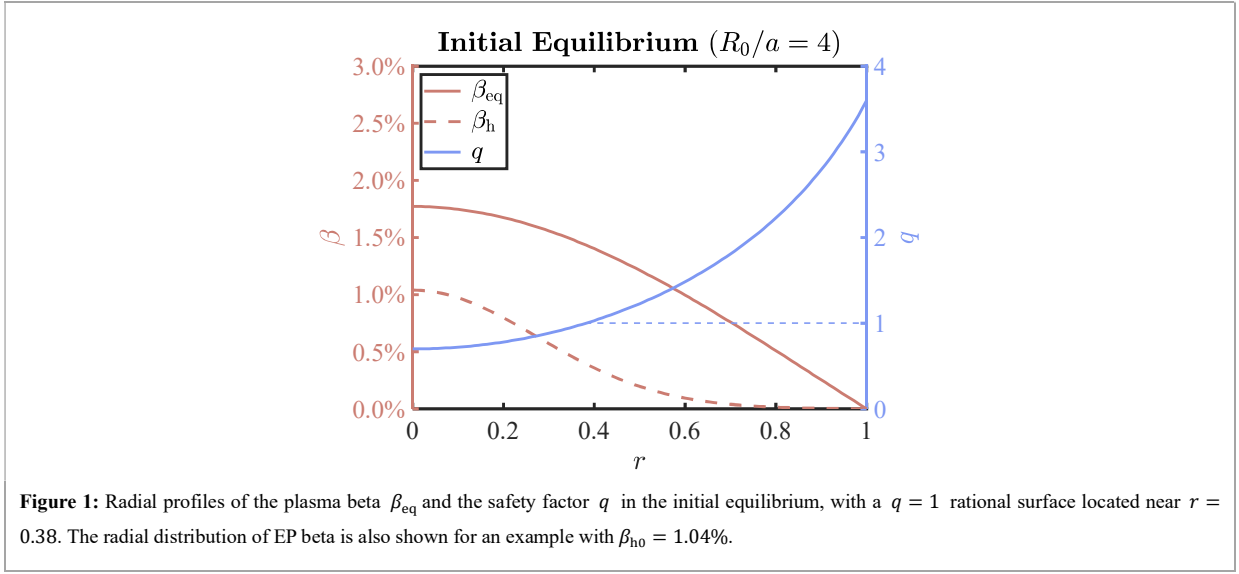

**Figure 1:** Radial profiles of the plasma beta $\beta_{eq}$ and the safety factor $q$ in the initial equilibrium, with a $q = 1$ rational surface located near $r = 0.38$. The radial distribution of EP beta is also shown for an example with $\beta_{h0} = 1.04\%$.

The dissipation coefficients (or their range) selected for this study are: $D = 2 \times 10^{-5}$, $\eta = (2 \sim 10) \times 10^{-6}$, $\nu = (1 \sim 2) \times 10^{-5}$, $\kappa_\perp = 1 \times 10^{-5}$, and $\kappa_\parallel = 1$. A uniform grid of $200 \times 32 \times 200$ was applied in the $\{R, \phi, Z\}$ directions, and $4{,}000{,}000$ non-uniform sampled markers (the sampling weight $g$ in the phase space is proportional to $f_0$) were used, with convergence analysis already performed. We employed multi-$n$ simulations, retaining all modes for $n = 0 \sim 7$ of both the MHD and EP components, and we will discuss the necessity of multi-$n$ simulations in Section 6.4.

## 4.  Pure MHD Simulation Results and Types of Sawtooth Oscillations

First, we consider the case without EP injection, i.e., the pure MHD simulation. In the chosen equilibrium and parameter space, the most unstable mode is the $1/1$ resistive IKM, as $q_0 = 0.7 < 1$. As expected, the IKM grows, saturates, and induces a sawtooth crash. Afterward, the system recovers under the influence of heat and current sources, leading to periodic sawtooth oscillations. We treat the resistivity and viscosity coefficients as tunable parameters, and after a prolonged simulation under different parameter settings, sawtooth oscillations may evolve into normal sawteeth, small sawteeth, or steady-island states, which we refer to as the three types of sawtooth oscillations. Figure 2 shows the time evolution of kinetic energy $\mathcal{E}_k$, radial magnetic field perturbations (measured by the imaginary part of Fourier components, $\mathrm{Im}(\delta \hat{B}_r)$, which indicates the size and phase of the magnetic island), the maximum value of radial flow speed (absolute value) $V_{r\max}$, core plasma pressure $p_{core}$, and the minimum safety factor $q_{\min}$ of the axisymmetric component [a] in three typical cases. Figure 3 presents Poincaré plots of magnetic field lines and flow patterns on $\phi = 0$ cross-section at specific moments in these three typical cases.

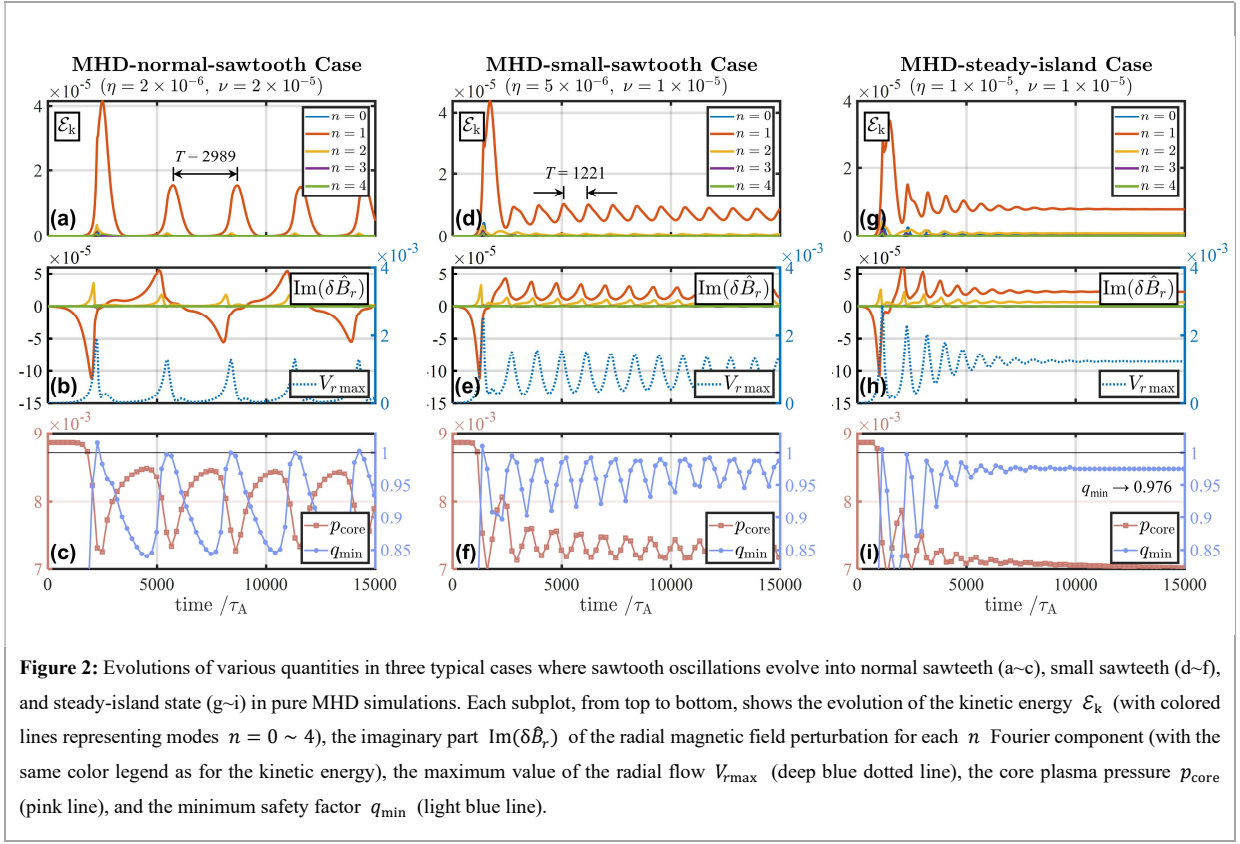

**Figure 2:** Evolutions of various quantities in three typical cases where sawtooth oscillations evolve into normal sawteeth (a~c), small sawteeth (d~f), and steady-island state (g~i) in pure MHD simulations. Each subplot, from top to bottom, shows the evolution of the kinetic energy $\mathcal{E}_k$ (with colored lines representing modes $n = 0 \sim 4$), the imaginary part $\mathrm{Im}(\delta \hat{B}_r)$ of the radial magnetic field perturbation for each $n$ Fourier component (with the same color legend as for the kinetic energy), the maximum value of the radial flow $V_{r\max}$ (deep blue dotted line), the core plasma pressure $p_{core}$ (pink line), and the minimum safety factor $q_{\min}$ (light blue line).

When $\eta = 2 \times 10^{-6}$ and $\nu = 2 \times 10^{-5}$, the system exhibits periodic normal sawtooth oscillations, as shown in Figure 2(a~c). Figure 3(a~c) show the evolution of the magnetic field structure before and after the first crash (precursor and post-cursor stages), illustrating that the $1/1$ magnetic island grows until it occupies the entire core and replaces the original magnetic axis, which is completely reconnected. This indicates a "complete" magnetic reconnection process similar to the Kadomtsev model [3], accompanied by a significant drop in plasma pressure, with $q \geq 1$ everywhere after the crash. After a duration of heating, the nested magnetic surfaces recover and a new magnetic island appears (as shown in Figure 3(c)), and the plasma pressure gradually recovers until the second crash. The magnetic field structure evolution during the second crash is shown in Figures 3(e~g), which is also driven by complete reconnection. From the evolution of the magnetic field structure, $\mathrm{Im}(\delta \hat{B}_r)$, and $q_{\min}$, it is evident that each newly formed magnetic island undergoes a phase reversal compared to the previous one (indicating the new magnetic axis appears at a different poloidal location), and the minimum safety factor after each crash reaches or slightly exceeds 1. When every sawtooth crash results from complete reconnection, the system exhibits normal sawtooth behavior. We refer to this case as the "MHD-normal-sawtooth Case". Interestingly, Figure 3(b) shows the formation of a broad stochastic magnetic field region outside the mixing radius after the first crash, which may be linked to the relatively large $|1 - q_0|$ value. The effect of stochastic fields will be discussed in Section 6.3.

---

[a] In this study, the method for calculating $q_{\min}$ involves performing Fourier filtering of the magnetic field, retaining only the $n = 0$ component (i.e., the axisymmetric component), then tracing magnetic field lines to compute the $q$ distribution across the entire space, with the minimum value taken as $q_{\min}$. This is done to avoid the influence of the $m/n = 1/1$ magnetic island on the $q_{\min}$ calculation, as the magnetic island always has $q \sim 1$ inside.

When $\eta = 5 \times 10^{-6}$ and $\nu = 1 \times 10^{-5}$, the system evolves differently from the normal sawtooth case. Although the first crash still involves complete magnetic reconnection (also related to the large $|1 - q_0|$ value), the system begins evolving into smaller amplitude, shorter period small sawtooth oscillations thereafter, as shown in Figure 2(d~f). Observing the Poincaré plot in Figure 3(i~k), it is evident that after transiting to the small sawtooth state, the $1/1$ magnetic island never occupies the entire core region. Instead, the size of the island undergoes periodic oscillations, and as a result, the phase of the island does not reverse, and the minimum safety factor always oscillates below 1 (as can also be seen from the evolution of $\mathrm{Im}(\delta \hat{B}_r)$ and $q_{\min}$). These features indicate that, in the small sawtooth state, the sawtooth crash behavior differs significantly different, more resembling "incomplete" or "partial" reconnection observed in many experimental devices [6,83-87] and in some simulations [18,19]. We refer to this case as the "MHD-small-sawtooth Case".

If the resistivity is further increased in this case, i.e., $\eta = 1 \times 10^{-5}$, the system will gradually transition into a quasi-steady state after a duration of small sawteeth, around $t = 8000\tau_A$, as shown in Figure 2(g~i). The kinetic energy, pressure, and $q_{\min}$ remain nearly constant, and the Poincaré plot reveals a steady $1/1$ magnetic island structure (Figure 3(m~o)), forming a helical steady magnetic field. This resembles the sawtooth-free advanced scenario often observed in many tokamak experiments [88-91]. We refer to this case as the "MHD-steady-island Case". In this case, the final $q_{\min}$ is less than 1, approximately 0.976.

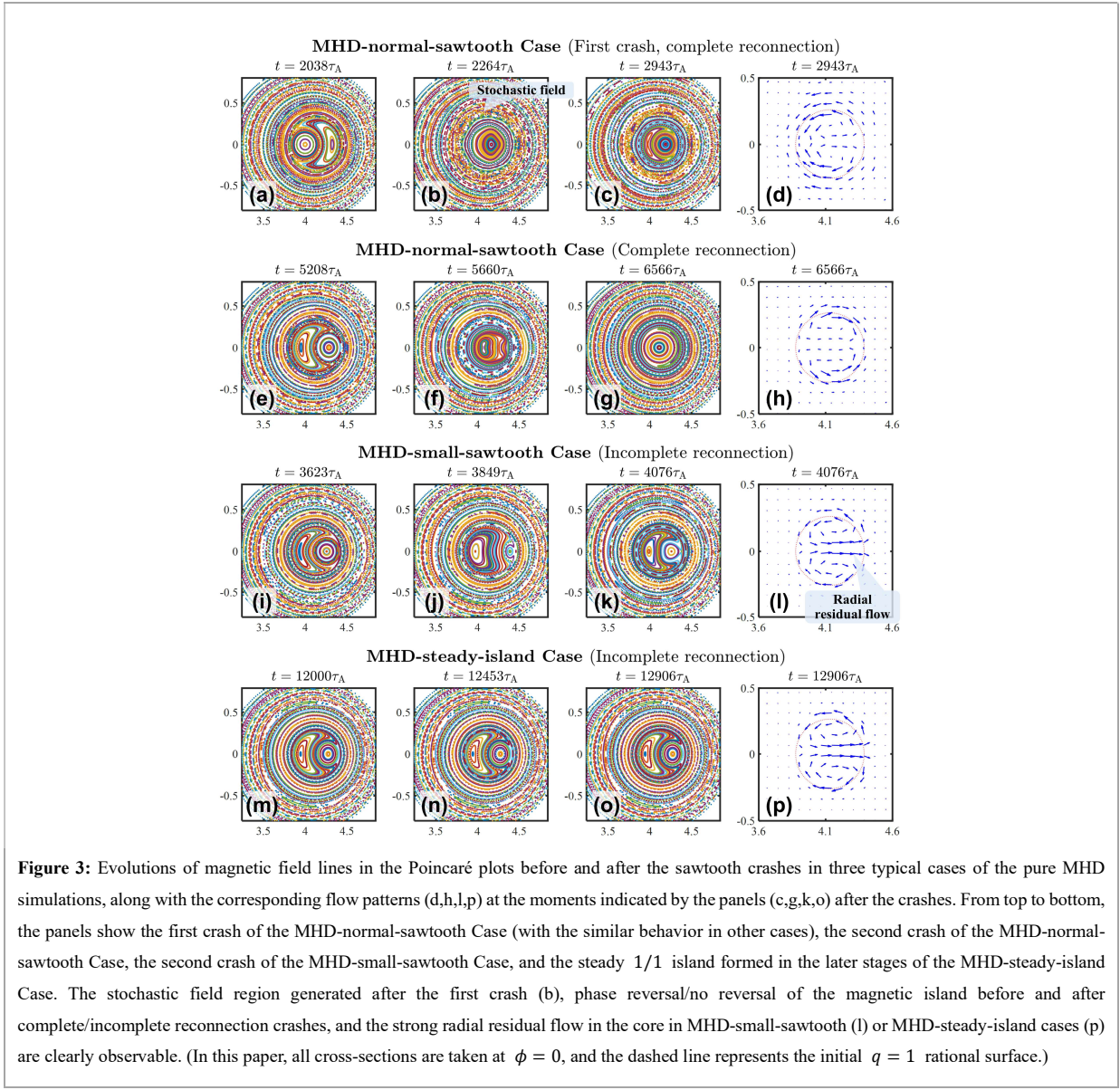

**Figure 3:** Evolutions of magnetic field lines in the Poincaré plots before and after the sawtooth crashes in three typical cases of the pure MHD simulations, along with the corresponding flow patterns (d,h,l,p) at the moments indicated by the panels (c,g,k,o) after the crashes. From top to bottom, the panels show the first crash of the MHD-normal-sawtooth Case (with the similar behavior in other cases), the second crash of the MHD-normal-sawtooth Case, the second crash of the MHD-small-sawtooth Case, and the steady $1/1$ island formed in the later stages of the MHD-steady-island Case. The stochastic field region generated after the first crash (b), phase reversal/no reversal of the magnetic island before and after complete/incomplete reconnection crashes, and the strong radial residual flow in the core in MHD-small-sawtooth (l) or MHD-steady-island cases (p) are clearly observable. (In this paper, all cross-sections are taken at $\phi = 0$, and the dashed line represents the initial $q = 1$ rational surface.)

Comparing the evolution of kinetic energy for the three cases (Figure 2(a,d,g)), we observe that in the normal sawtooth case, the kinetic energy nearly drops to zero after each sawtooth cycle. However, in the small sawtooth and steady-island cases, significant residual kinetic energy remains after each sawtooth cycle. This residual energy is related to the radial residual flow in the core, as shown by the evolution of $V_{r\mathrm{ma}}$ in Figure 2(b,e,h), where strong residual flow can be seen in panels (e) and (h). Previous studies [18,30] have shown that the presence of radial residual flow is crucial for the system's evolution into a small sawtooth or steady-state, as it continually transports magnetic flux and energy out of the plasma core, preventing further increases in core current density and pressure. As seen in Figure 2(c,f,i), the pressure in the core after each sawtooth cycle in panels (f) and (i) settles at a much lower level than in panel (c). This prevents the accumulated thermal energy in the core from reaching levels sufficient to drive normal sawteeth, leading the system to evolve into small oscillations or even a steady-state. In Figure 3(d,h,l,p), we plot the flow patterns corresponding to post-cursor moments after the crash, showing that the radial residual flow is much stronger in the small sawtooth and steady-island cases compared to the normal sawtooth case. This mechanism is referred to as magnetic flux pumping. While early works [16,17,66] focused on $1/1$ quasi-interchange modes to achieve $q \geq 1$, thereby completely preventing sawtooth oscillations, it is also applicable to the analysis of the $1/1$ IKM with $q < 1$ here.

For clarity, we summarize the main characteristics of the three types of sawtooth oscillations in Table 1. In addition to the Poincaré plot of magnetic field lines, whether $q_{min}$ exceeds 1 after the crash can also serve as a criterion for determining the type of reconnection and sawteeth.

**Table 1:** Comparison of the characteristics of the three types of sawtooth oscillations.

| Sawtooth Type | Normal Sawtooth | Small Sawtooth | Steady Island State |
|---|---|---|---|
| Amplitude | Large | Small | No oscillation |
| Period | Long | Short | No oscillation |
| Magnetic Axis Disappearance | Yes | No | No |
| $1/1$ Magnetic Island Phase | May reverse | Does not reverse | Does not reverse |
| Reconnection Type | Complete | Incomplete | Incomplete |
| Core Radial Residual Flow | Small | Large | Large |
| $q_{min}$ after Crash | $> 1$ | $< 1$ | $< 1$ |
| Typical Parameters | $\eta = 2 \times 10^{-6}$ $\nu = 2 \times 10^{-5}$ | $\eta = 5 \times 10^{-6}$ $\nu = 1 \times 10^{-5}$ | $\eta = 1 \times 10^{-5}$ $\nu = 1 \times 10^{-5}$ |
| Corresponding Case Name | MHD-normal-sawtooth Case | MHD-small-sawtooth Case | MHD-steady-island Case |

In the above discussion, we have shown that resistivity $\eta$ and viscosity $\nu$ significantly affect the type of long-term sawtooth behavior. The results from scanning a larger parameter space are presented in Figure 4. At a low resistivity, a moderate viscosity leads to normal sawteeth, while both lower and higher viscosities tend to result in small sawteeth or even steady-states. The steady-state at a very high viscosity agrees with Shen and Porcelli's results [19]. The steady-state at a lower viscosity is consistent with Zhang et al.'s findings [18], as the low viscosity is insufficient to dissipate the residual radial flow in the core. At a higher resistivity, the system is more likely to evolve to a steady-state, which has been less emphasized in previous studies. The figure also includes isolines of the magnetic Prandtl number $\mathcal{P}$, defined as $\mathcal{P} = \nu/\eta$. For computational feasibility, the values of $\nu$ and $\eta$ used in our simulations are several orders of magnitude higher than those in experiments. However, the typical cases used in this paper (marked in the figure) have $\mathcal{P}$ in the range of $1 \sim 10$, which are comparable to experimental devices.

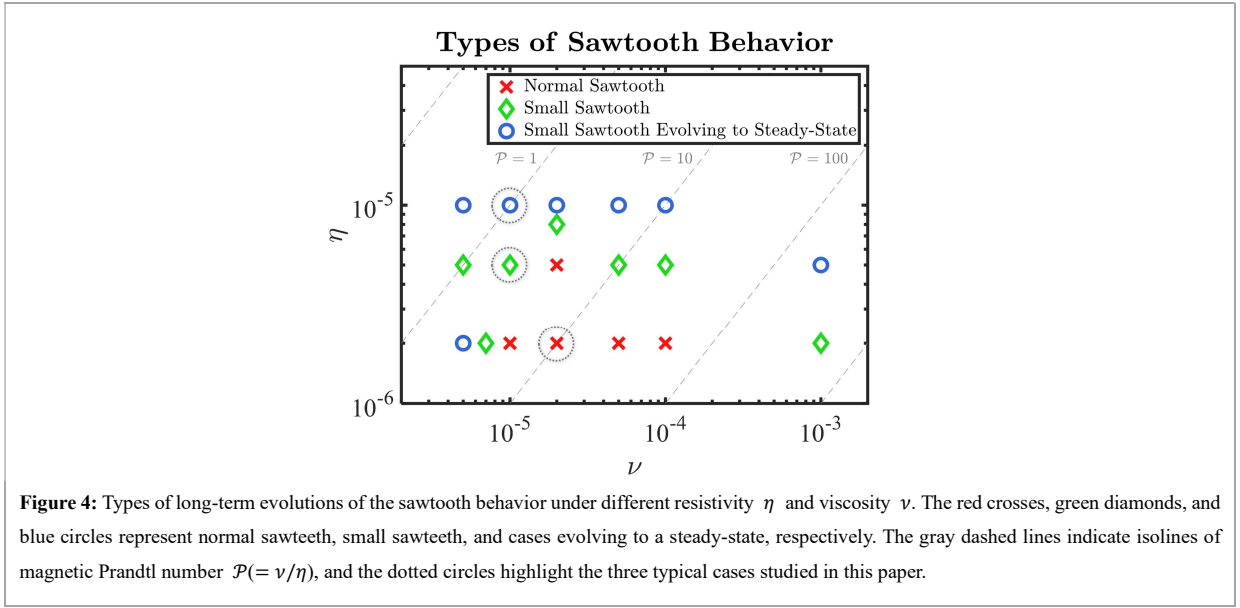

**Figure 4:** Types of long-term evolutions of the sawtooth behavior under different resistivity $\eta$ and viscosity $\nu$. The red crosses, green diamonds, and blue circles represent normal sawteeth, small sawteeth, and cases evolving to a steady-state, respectively. The gray dashed lines indicate isolines of magnetic Prandtl number $\mathcal{P}(=\nu/\eta)$, and the dotted circles highlight the three typical cases studied in this paper.

## 5. Injection of Passing EPs: Simulation Results and Physical Interpretation

### 5.1 Simulation Results of EP Injection

In this section, we inject passing EPs into the three typical MHD cases from the previous section and study their influences on sawtooth types. We first analyze the MHD-steady-island Case as an example, as its phenomena are the most prominent.

In Reference [69], we discussed the sensitivity of the $1/1$ IKM simulation results to the model and pointed out that for EPs, in principle, both the $\delta f$ contribution and $\delta\boldsymbol{B}$ contribution should be considered. When considering the $\delta\boldsymbol{B}$ contribution, the EP pressure must be subtracted from the initial equilibrium pressure. However, as discussed in Reference [69], considering that the EP pressure is an anisotropic tensor and is not a function of the magnetic surface (in the slowing-down distribution, it is dependence on $\langle\psi\rangle$ rather than $\psi$), self-consistently calculating the bulk pressure after subtracting the EP pressure is difficult and may lead to inaccuracies due to mechanical imbalances. Additionally, changing the bulk pressure profile shape would complicate comparisons between cases. Fortunately, as we found in Reference [69], for passing EPs, the effect of the $\delta\boldsymbol{B}$ contribution is minimal. Therefore, to simplify the analysis, most simulations in this paper only consider the $\delta f$ contribution of EPs (i.e., "Scheme A" in Reference [69]) and assuming that the equilibrium pressure is only provided by the bulk plasma and treating the EP pressure as an external addition. To present a more comprehensive consideration, in Section 5.4, we also show results using a more self-consistent scheme, including both the $\delta f$ and $\delta\boldsymbol{B}$ contributions of EPs (i.e., "Scheme B" in Reference [69], with EP pressure subtracted from the equilibrium pressure). Our analysis confirms that these two different choices do not affect the main conclusions discussed in this paper.

Based on the MHD-steady-island Case, passing EPs are injected in both co-current and counter-current directions, with EP beta values of $1.04\%$ (high fraction) and $0.52\%$ (low fraction). The evolutions of the kinetic energy and the magnetic field structure in the later stage are shown in Figure 5.

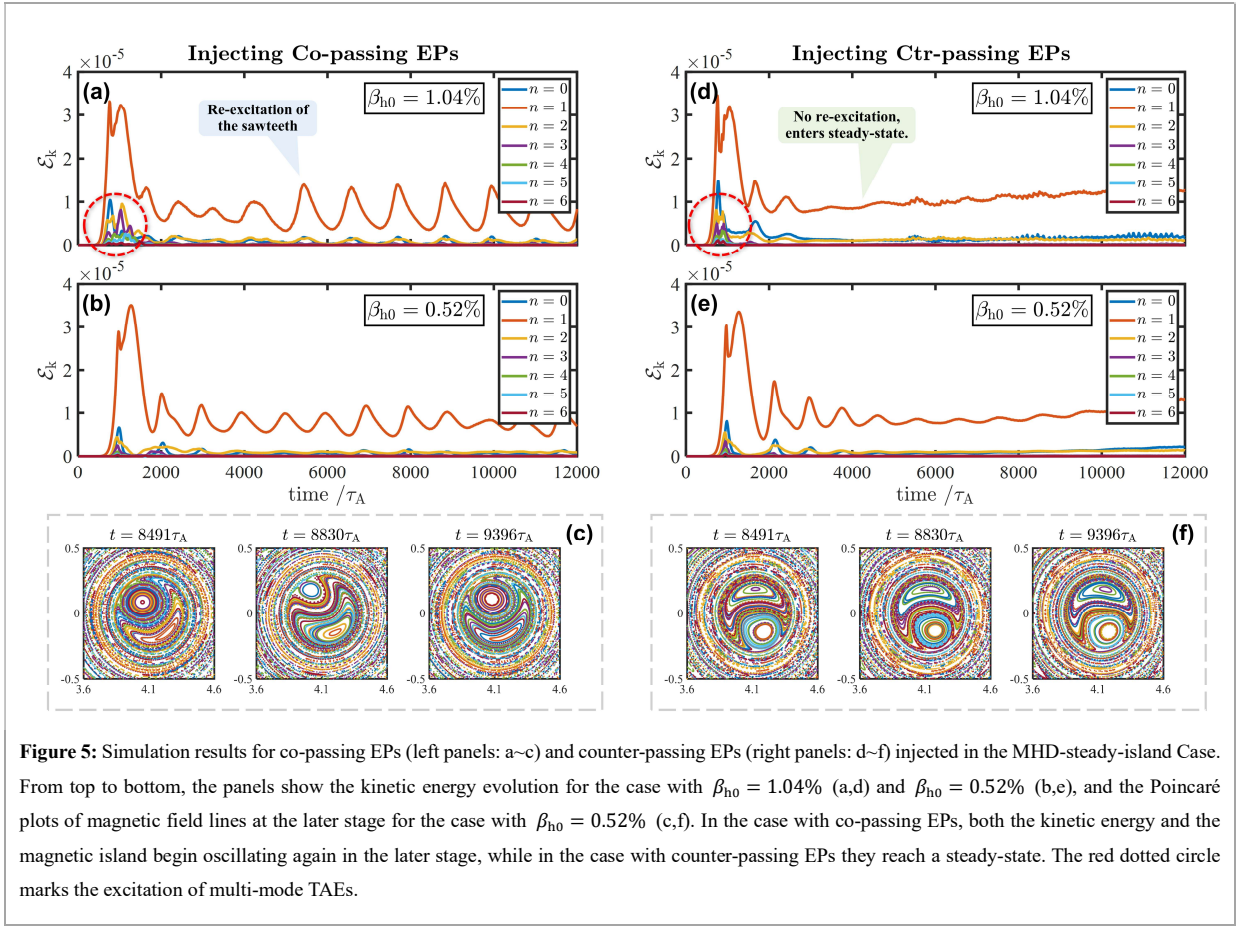

**Figure 5:** Simulation results for co-passing EPs (left panels: a~c) and counter-passing EPs (right panels: d~f) injected in the MHD-steady-island Case. From top to bottom, the panels show the kinetic energy evolution for the case with $\beta_{h0} = 1.04\%$ (a,d) and $\beta_{h0} = 0.52\%$ (b,e), and the Poincaré plots of magnetic field lines at the later stage for the case with $\beta_{h0} = 0.52\%$ (c,f). In the case with co-passing EPs, both the kinetic energy and the magnetic island begin oscillating again in the later stage, while in the case with counter-passing EPs they reach a steady-state. The red dotted circle marks the excitation of multi-mode TAEs.

In Figure 5, we observe that when co-passing EPs are injected, the system does not reach a steady-state as in the pure MHD simulation, but instead, after some time, new periodic sawtooth oscillations are re-excited and maintain a certain amplitude. This is referred to as "re-excitation" of the sawtooth by co-passing EPs. In contrast, when counter-passing EPs are injected, no sawtooth re-excitation occurs, and the system reaches a steady-state, forming a nearly static $1/1$ magnetic island. This demonstrates a significant difference in the impact of co-passing and counter-passing EPs on the sawtooth type. Interestingly, when counter-passing EPs are injected, after the system reaches a steady-state, kinetic energy increases again in the later stages of evolution. However, this does not occur with co-passing EPs. This phenomenon will be discussed in Section 6.2 of this paper.

Changing the EP fraction does not affect the qualitative conclusions but influences the amplitude of the sawtooth after re-excitation (for co-passing EPs) and the time to reach steady-state (for counter-passing EPs). Furthermore, when higher fractions of EPs ($\beta_{h0} = 1.04\%$) are injected, we observe more pronounced excitation of multiple high-$n$ modes after the first crash. These high-$n$ modes are identified as global TAEs excited by the sawtooth crash, which will be analyzed in detail in Section 6.1 of this paper.

## 5.2 Redistribution of EPs Caused by Sawtooth Crash

Sawtooth crash leads to significant transport of EPs, resulting in their strong redistribution. Consider the simplest case: when EP energy is not too high, EPs can be assumed to move along the magnetic field lines. Initially, EPs are peaked at the magnetic axis in the core. As the IKM develops, the $1/1$ magnetic island grows, pushing the magnetic axis outward, causing the peak of the EP distribution to shift outward as well. When the sawtooth crash with the complete reconnection occurs, the original magnetic axis disappears, and the magnetic surfaces around the magnetic axis are also destroyed. These EPs with relatively lower energy escape from the core region into areas outside the $q = 1$ surface, where they are further randomly distributed by the broad stochastic field present in that region.

Since all three typical cases involve a complete reconnection during the first sawtooth crash, all cases involve EPs experiencing this process. Figure 6 shows a typical example, using $P_{h\parallel}$ distribution to measure EP distribution (since for passing EPs we are studying, $P_{h\parallel} \gg P_{h\perp}$). The direction of EP escape aligns with the orbit shift direction (co-passing EPs shift toward the weak field side, counter-passing EPs shift toward the strong field side). Ultimately, EPs accumulate in a ring-like region outside the $q = 1$ surface, with a nearly symmetric distribution in both toroidal and poloidal directions (indicating that $\delta \mathbf{P_h}$ shifts from $n = 1$ dominance in the linear stage to $n = 0$ dominance in the nonlinear stage). Since EPs accumulated in the outer region are dissipated quickly with time, thus this redistribution does not reverse the pressure gradient direction at $q = 1$, despite significant changes in the EP pressure profile as seen in Figure 6. Therefore, after the crash, EPs remain "on-axis" rather than "off-axis".

Many previous theoretical and simulation studies [39-44] have shown that when EP energy exceeds a critical threshold, their orbits tend to decouple from the IKM, leading to weaker redistribution. The level of EP redistribution depends not only on energy but also on pitch angle, with co-passing EPs generally more susceptible to redistribution than counter-passing or trapped EPs [43-47]. In our results, this trend is reflected in Figure 6, where the redistribution of co-passing EPs is slightly more pronounced than that of counter-passing EPs. It should be noted that in our equilibrium, the relatively large value of $|1 - q_0|$ leads to the appearance of stochastic magnetic fields around the $1/1$ magnetic island before the crash, and the stochastic fields spread over a board region outside the initial $q = 1$ surface after the crash (as clearly shown in Figure 6), further affecting EP transport. Preliminary discussions on this will be provided in Section 6.3.

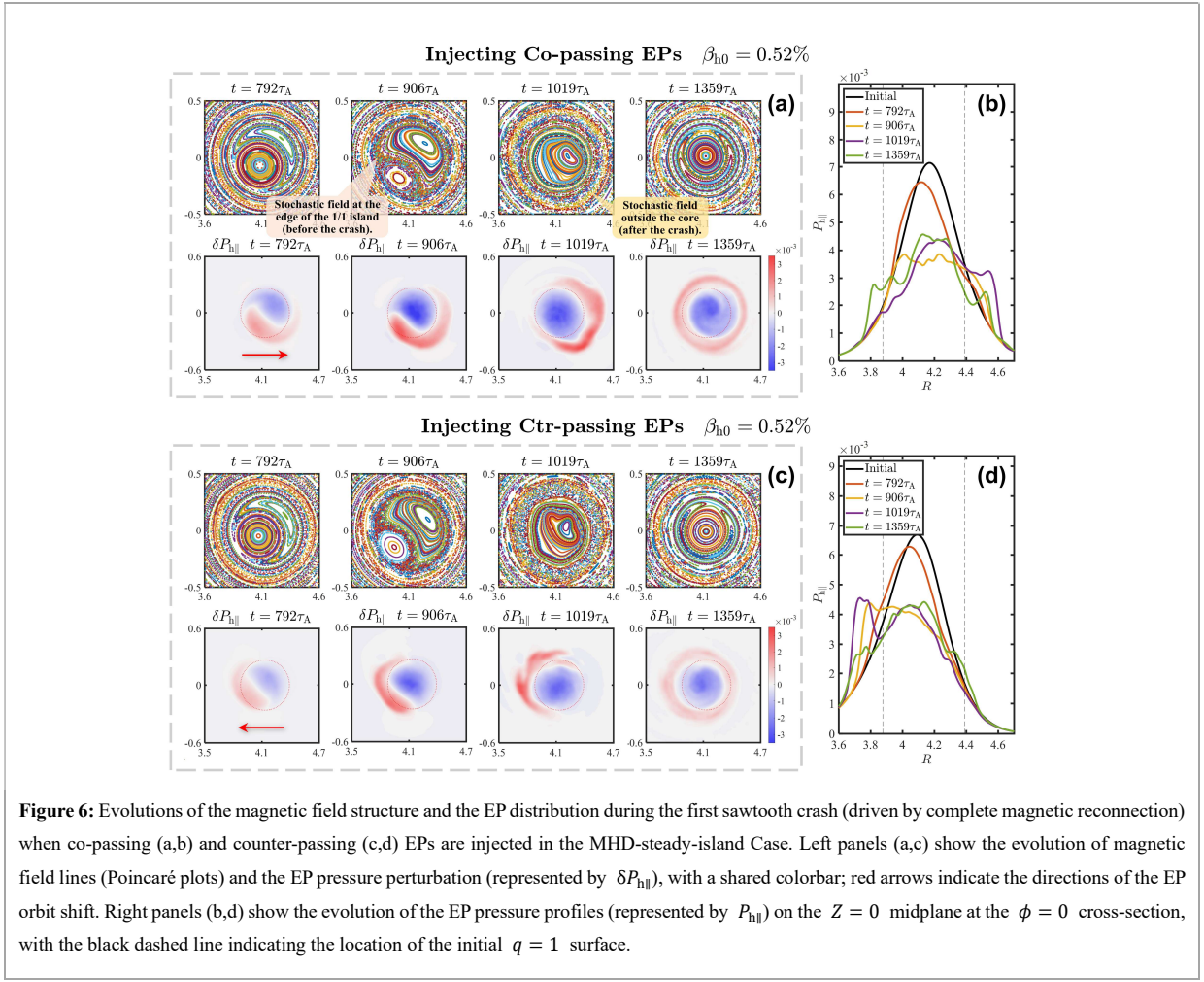

**Figure 6:** Evolutions of the magnetic field structure and the EP distribution during the first sawtooth crash (driven by complete magnetic reconnection) when co-passing (a,b) and counter-passing (c,d) EPs are injected in the MHD-steady-island Case. Left panels (a,c) show the evolution of magnetic field lines (Poincaré plots) and the EP pressure perturbation (represented by $\delta P_{h\parallel}$), with a shared colorbar; red arrows indicate the directions of the EP orbit shift. Right panels (b,d) show the evolution of the EP pressure profiles (represented by $P_{h\parallel}$) on the $Z = 0$ midplane at the $\phi = 0$ cross-section, with the black dashed line indicating the location of the initial $q = 1$ surface.

## 5.3 Physical Picture of EPs' Impact on Sawtooth Oscillation Types

The main objective of this paper is to discuss the impact of EPs on the transition of sawtooth oscillation types from the perspective of nonlinear evolution. To clearly explain this process, we focus on the evolution of a series of physical quantities over time. We compare the cases of injecting co-passing EPs and counter-passing EPs, as shown in Figure 7. All panels use purple-red lines to represent the case of co-passing EP injection, and blue lines to represent the case of counter-passing EP injection.

The evolution of kinetic energy ($n = 1$) and the corresponding instantaneous growth rate $\gamma$ for different EP injection directions are shown in Figures 7(a) and 7(b), respectively. We focus on the changes of sawtooth periods during the early nonlinear stage of several sawtooth cycles. When co-passing EPs are injected, the sawtooth period tends to gradually increase after the first crash, while with counter-passing EPs, it decreases. Previous theoretical studies [55-58,60,61] have shown that counter-passing EPs on-axis have a relatively destabilizing effect on the IKM compared to co-passing EPs, which can lead to higher instantaneous energy growth rates and consequently shorten the sawtooth period. This simulation result is consistent with theoretical predictions. As can be seen in Figure 7(b), in the early nonlinear stage, although the sawtooth phase in the case with counter-passing EPs initially lags behind, the higher instantaneous energy growth rate plays a crucial role in altering the sawtooth period, causing the phase to quickly advance ahead of the case with co-passing EPs.

Changes in the sawtooth period can affect the residual level of the core radial flow after each sawtooth cycle. Figure 7(c) compares the evolution of $V_{r\max}$ for both cases. During each cycle, the core radial flow is always dissipated by the viscosity, and a shorter sawtooth period typically means that this flow is not fully dissipated within a single cycle, leaving a residual that continuously pumps heat and flux from the core to the reconnection region, helping maintain a steady-island state. Injecting co-passing EPs into the system, which would otherwise evolve into a steady-island state in a pure MHD simulation, causes the sawtooth period to lengthen. After several cycles, the period becomes long enough for the core flow to dissipate more, and when the residual flow level drops below a certain threshold, it can no longer move the heat and flux to the reconnection region. As a result, the steady-state can no longer be maintained, and sawteeth are re-excited. This mechanism is similar to the principle observed in previous studies, where ECCD deposited outside the $q = 1$ surface leads to the re-excitation of sawtooth oscillations in originally non-sawtooth cases (see Figure 11 in Reference [30]). The threshold is approximately indicated by the gray dashed line in Figure 7(c). It is important to note that due to EPs, the distortion of mode structures means that $V_{r\max}$ can only qualitatively reflect the strength of the radial residual flow, and thus this threshold is an estimate.

To more intuitively demonstrate the role of core radial residual flow on reconnection, we plot the evolution of the toroidal electric field $E_\phi$ at the X-point [b] in Figure 7(d), which reflects the reconnection rate (stronger X-point toroidal electric field $E_{\phi,X}$ indicates a higher reconnection rate; note that $E_{\phi,X}$ is negative due to the reversed electric field, and here we refer to its absolute value). Figure 7(e) shows the evolution of the minimum safety factor $q_{\min}$. By comparing the trends of $V_{r\max}$, $E_{\phi,X}$, and $q_{\min}$ within the same sawtooth cycle, it is clear that the levels of the residual flow, the reconnection rate, and $q_{\min}$ change synchronously. Stronger residual flow maintains higher reconnection rates near the X-point, preventing further decrease of $q_{\min}$ and making the system more likely to transition into a steady-island state without sawtooth oscillations.

---

[b]    The method we used to determine the locations of the X-point (and O-point) is the Poincaré-Hopf index method, based on the zero-point theorem of vector fields [92]. The detailed steps can be found in Reference [93]. Since we only consider a "once around" mapping, only the magnetic axis and the O-point and X-point related to the $1/1$ rational surface are marked in Figure 8. If multiple X-points are present, the primary X-point can be tracked.

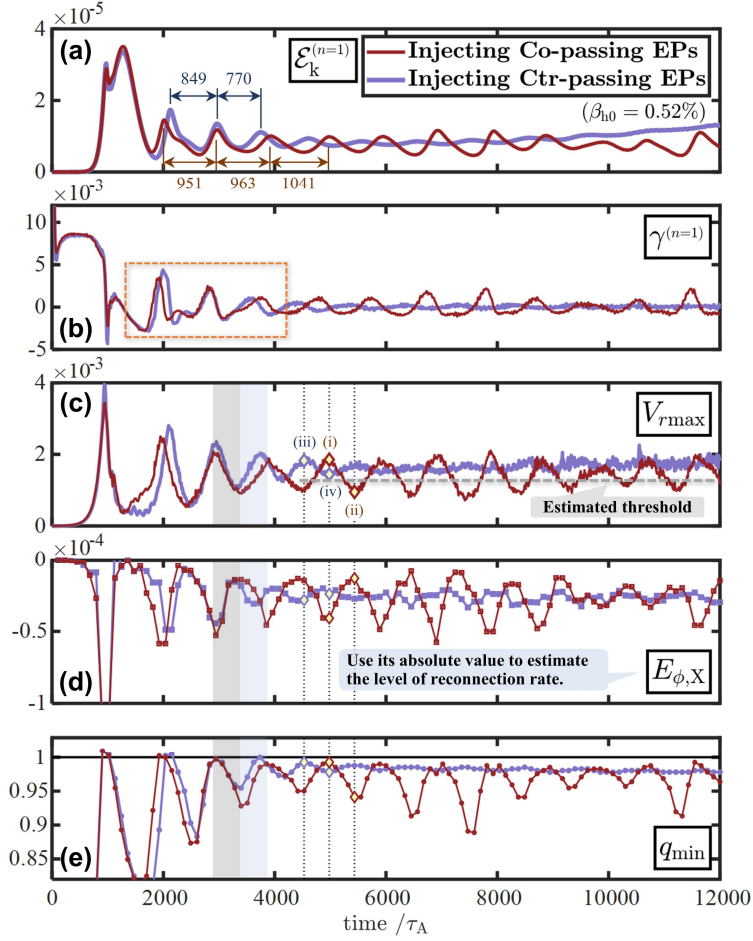

**Figure 7:** Evolutions of the kinetic energy $\mathcal{E}_k$ ($n = 1$ component), the instantaneous growth rate of the kinetic energy $\gamma$ ($n = 1$ component), the maximum radial flow $V_{r\max}$, the toroidal electric field at the X-point $E_{\phi,X}$ (used to estimate the reconnection rate through its absolute value), and the minimum safety factor $q_{\min}$ when passing EPs ($\beta_{h0} = 0.52\%$) are injected in the MHD-steady-island Case. In all panels, red lines represent the injection of co-passing EPs, while blue lines represent the injection of counter-passing EPs. The orange dashed box highlights the early nonlinear stage, and the gray dashed line indicates the estimated threshold of radial residual flow. Light red and light blue shaded areas are time-aligned to assist in comparing the trends of the three physical quantities within a single sawtooth cycle. Diamonds and black dotted lines mark the moments observed in Figure 8.

We select two moments when the radial flow is strongest and weakest within a sawtooth cycle from the co-passing EP case (i) $t = 4981\tau_A$ and (ii) $t = 5434\tau_A$ and two similar moments from the counter-passing EP case (iii) $t = 4528\tau_A$ and (iv) $t = 4981\tau_A$ (marked in Figure 7), with the corresponding poloidal cross-sectional flow patterns and magnetic field line Poincaré plots shown in Figure 8 (the O-point and X-point positions are also marked). In the case of the co-passing EP injection, radial residual flow is significantly reduced after the sawtooth cycle, while in the counter-passing EP case, this reduction is negligible. From the perspective of the magnetic flux pumping [18,30,66,94,95]: in our model, the magnetic flux source (the '$-\eta J_0$' term in Ohm's law) continually pumps magnetic flux into the core region, and the radial flow directs toward the X-point, thereby transporting the flux to the reconnection region for annihilation. Therefore, the magnitude of the radial flow determines the flux annihilation rate (i.e., the reconnection rate). When a strong radial flow carries away too much flux from the core, it is no longer sufficient to drive sawtooth oscillations, and the system maintains a steady-state. Conversely, if the radial flow is reduced, the sawtooth may be re-excited.

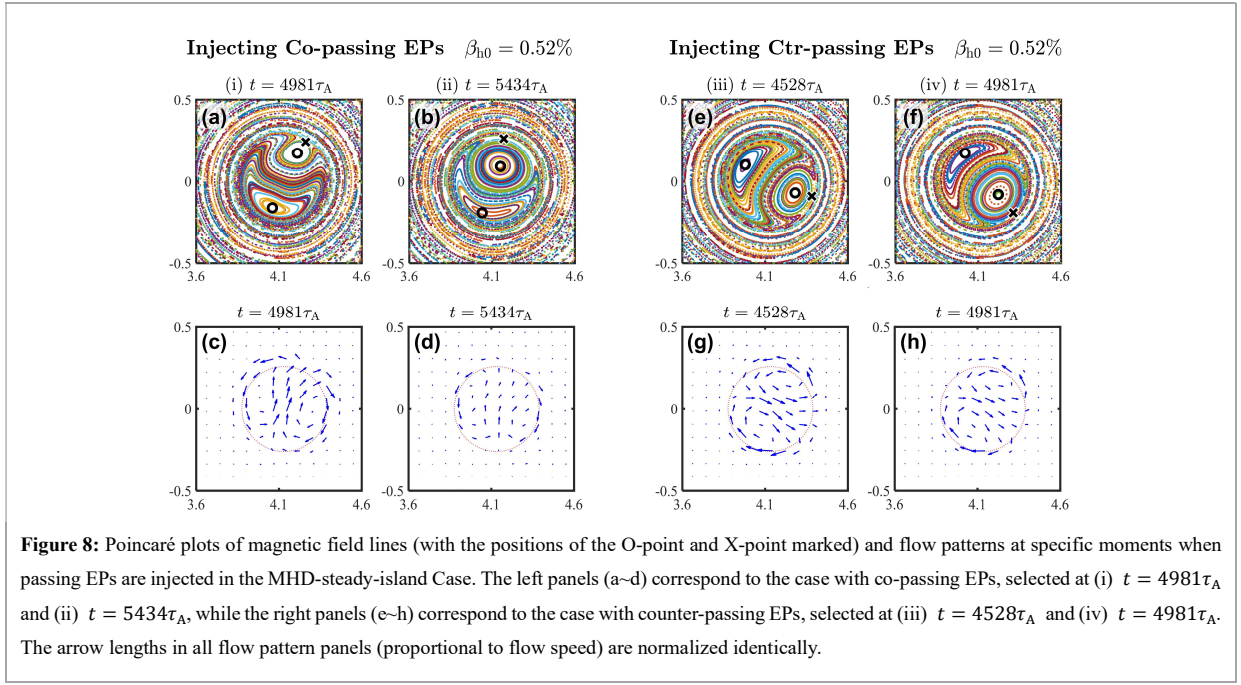

**Figure 8:** Poincaré plots of magnetic field lines (with the positions of the O-point and X-point marked) and flow patterns at specific moments when passing EPs are injected in the MHD-steady-island Case. The left panels (a–d) correspond to the case with co-passing EPs, selected at (i) $t = 4981\tau_A$ and (ii) $t = 5434\tau_A$, while the right panels (e–h) correspond to the case with counter-passing EPs, selected at (iii) $t = 4528\tau_A$ and (iv) $t = 4981\tau_A$. The arrow lengths in all flow pattern panels (proportional to flow speed) are normalized identically.

In summary, we outline the physical picture of how passing EPs affect sawtooth types: When co-passing EPs are injected, their stabilizing effect on the IKM extends the sawtooth period, reducing the radial residual flow level at the end of each sawtooth cycle. This leads to lower reconnection rates near the X-point, causing heat and magnetic flux to accumulate in the core, which tends to result in larger-amplitude sawtooth oscillations. In contrast, when counter-passing EPs are injected, their destabilizing effect on the IKM shortens the sawtooth period, resulting in stronger radial residual flow and higher reconnection rates near the X-point, favoring the formation of a steady-state magnetic island. The results of long-term nonlinear simulations are consistent with both theoretical predictions [55-58] and experimental observations [36,50-52].

### 5.4 Final Simulation Results: Control of Sawtooth Type by Passing EPs

Finally, for a more general case, this section presents the simulation results of injecting co-passing EPs and counter-passing EPs in three different parameter sets: the MHD-normal-sawtooth Case, the MHD-small-sawtooth Case, and the MHD-steady-island Case. To allow for a more self-consistent comparison, these cases in this section use a more complete coupling scheme, including the $\delta\boldsymbol{B}$ contribution from EPs and subtracting the EP pressure from the equilibrium pressure (i.e., "Scheme B" in Reference [69]). To minimize the impact of EPs on the bulk plasma pressure profile shape and reduce errors caused by the anisotropic pressure of EPs, we lower the EP pressure to $\beta_{h0} = 0.37\%$. The evolution of the kinetic energy (the $n = 1$ component), the maximum radial flow $V_{r\mathrm{max}}$, and $q_{\mathrm{min}}$ for all cases are shown in Figure 9.

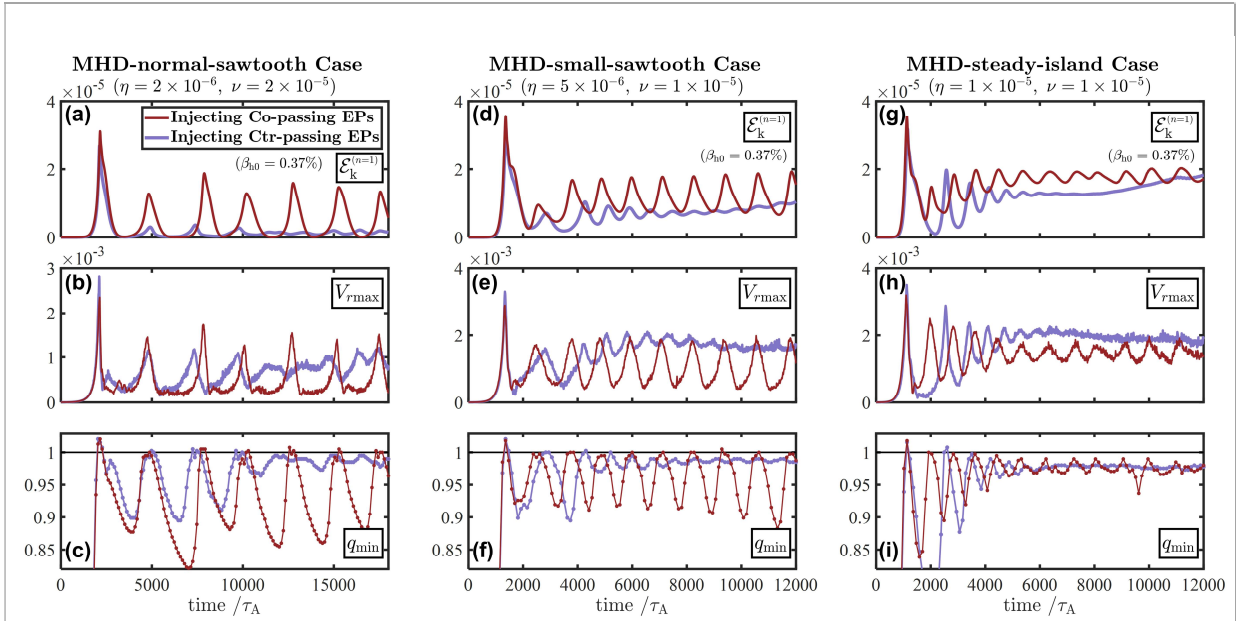

**Figure 9:** Evolutions of the kinetic energy $\mathcal{E}_k$ ($n = 1$ component), the maximum radial flow $V_{rmax}$, and the minimum safety factor $q_{min}$ when passing EPs ($\beta_{h0} = 0.37\%$) are injected in the three cases with different parameters. The $\delta\boldsymbol{B}$ contribution of EPs is included, and the EP pressure is subtracted from the equilibrium pressure. In all panels, red lines represent the injection of co-passing EPs, while blue lines represent the injection of counter-passing EPs. For the MHD-normal-sawtooth Case (a~c), injecting co-passing EPs maintains large-amplitude normal sawteeth, while counter-passing EPs lead to small sawteeth. For the MHD-small-sawtooth Case (d~f), co-passing EPs increase amplitude and period, while counter-passing EPs reduce both, potentially reaching a steady-state. For the MHD-steady-island Case (g~i), counter-passing EPs maintain the steady-state, while co-passing EPs re-excite small sawteeth. Cases with counter-passing EPs show higher levels of radial residual flow.

Overall, co-passing EPs tend to excite sawtooth oscillations, increasing their period and amplitude. In contrast, counter-passing EPs tend to transform normal sawteeth into small sawteeth, shortening their period and reducing their amplitude, and may even fully suppress them, leading the system to evolve into a steady-state. With the same parameters, injecting co-passing EPs makes complete reconnection (with $q_{min} > 1$ after the crash) more likely, while injecting counter-passing EPs favors incomplete reconnection (with $q_{min} < 1$ after the crash), where stronger radial residual flows are typically present. These conclusions are qualitatively consistent with our previous analysis and further demonstrate that different considerations of the coupling scheme do not affect the main physical picture.

# 6. Further Discoveries

## 6.1 Global Multi-$n$ TAEs Caused by Sawtooth Crash

In Section 5.1, we noted that when the injected EP fraction is high enough, a short-term enhancement of multiple high-$n$ components in the kinetic energy occurs after the first sawtooth crash (as highlighted by red circles in Figure 5(a,d)), indicating the excitation of multi-$n$ modes in the system. Figure 10 shows the mode structure (represented by the toroidal electric field $E_\phi$) at the moment before and after the first sawtooth crash in a typical case. It can be observed that these high-$n$ modes appear outside the $q = 1$ rational surface after the first crash and form a global mode with a broad distribution.

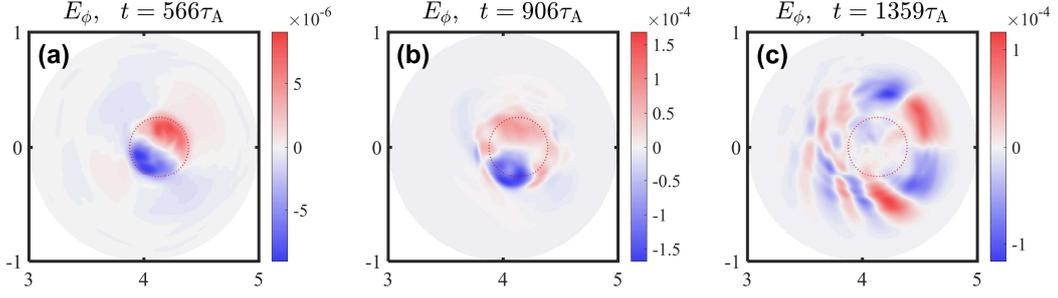

**Figure 10:** The toroidal electric field $E_\phi$ distribution before ($t = 566\tau_A$), during ($t = 906\tau_A$), and after ($t = 1359\tau_A$) the first sawtooth crash in the MHD-steady-island Case with co-passing EPs ($\beta_{h0} = 0.62\%$) injected. Before the first crash, the mode structure is dominated by the core $1/1$ IKM. Starting from the crash moment, high-$n$ global TAEs appear outside the $q = 1$ surface and grow.

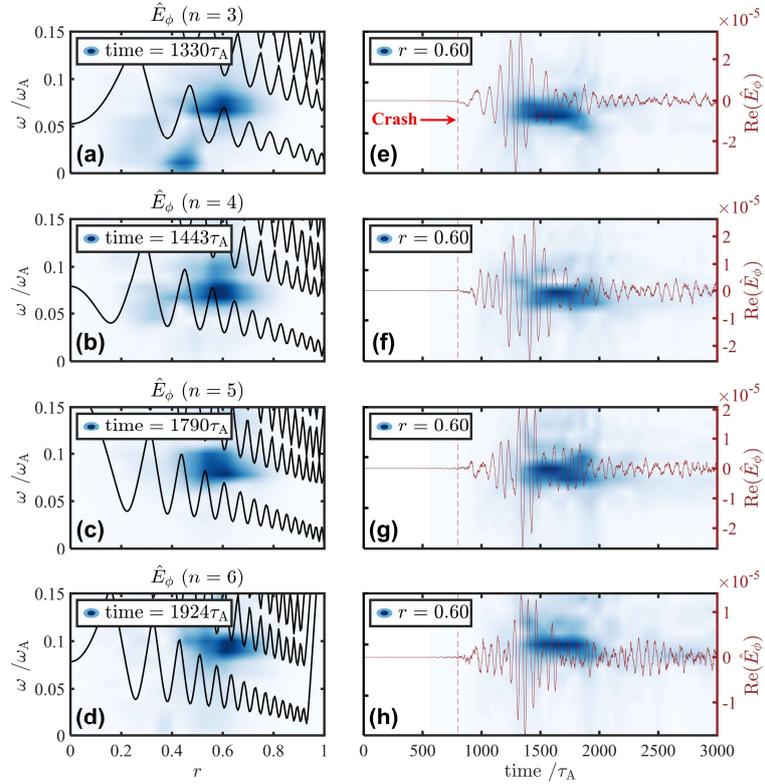

**Figure 11:** Spectra of the toroidal electric field Fourier components $\hat{E}_\phi$ for each $n$ mode (from top to bottom: $n = 3, 4, 5, 6$) with co-passing EPs ($\beta_{h0} = 0.62\%$) injected in the MHD-steady-island Case. The left panels show the radial spatial distribution of the spectrum at specific moments, with the black curve representing the continuous spectrum calculated by the NOVA code. The right panels show the time evolution of the spectrum at $r = 0.60$, with the dark red curve representing the real part of $\hat{E}_\phi$, and the red dashed line indicating the moment of the first sawtooth crash.

These modes are identified as global TAEs, as their frequencies are located in the "TAE gap" of the shear Alfvén wave continuous spectrum (shown in Figure 11). Over time, these modes show a downward frequency sweep and a spatial outward shift. The excitation of multi-$n$ TAEs indicate strong multi-mode nonlinear interactions in the system, highlighting the necessity of multi-$n$ nonlinear simulations.

The excitation of multi-mode TAEs requires a sufficiently strong radial gradient drive from EPs, meaning that the $\beta_{h0}$ must exceed a certain threshold. We found that sawteeth significantly promote the excitation of multi-mode TAEs, lowering the threshold for the required $\beta_{h0}$. To verify this, we fixed $\beta_{h0} = 0.62\%$ and simulated the system evolution by retaining all $n$ modes and filtering out the $n = 1$ mode, respectively. In the former case, a sawtooth crash occurs, while in the latter, no sawtooth behavior is observed. The evolution of the EP pressure and the amplitudes of the different $n$-modes for both cases are shown in Figure 12, with the same axes for comparison. The strong EP redistribution caused by the sawtooth crash increases the EP pressure gradient outside $q = 1$, exciting multi-mode TAEs. This excitation results from enhanced free energy in the outer region due to EP redistribution. In contrast, without the sawtooth crash, the EPs distribution remains nearly unchanged, and no TAE excitation is observed, with high-$n$ modes remaining at noise levels. Thus, the $n = 1$ mode associated with the sawtooth serve as a "seed" for the multi-mode TAE excitation.

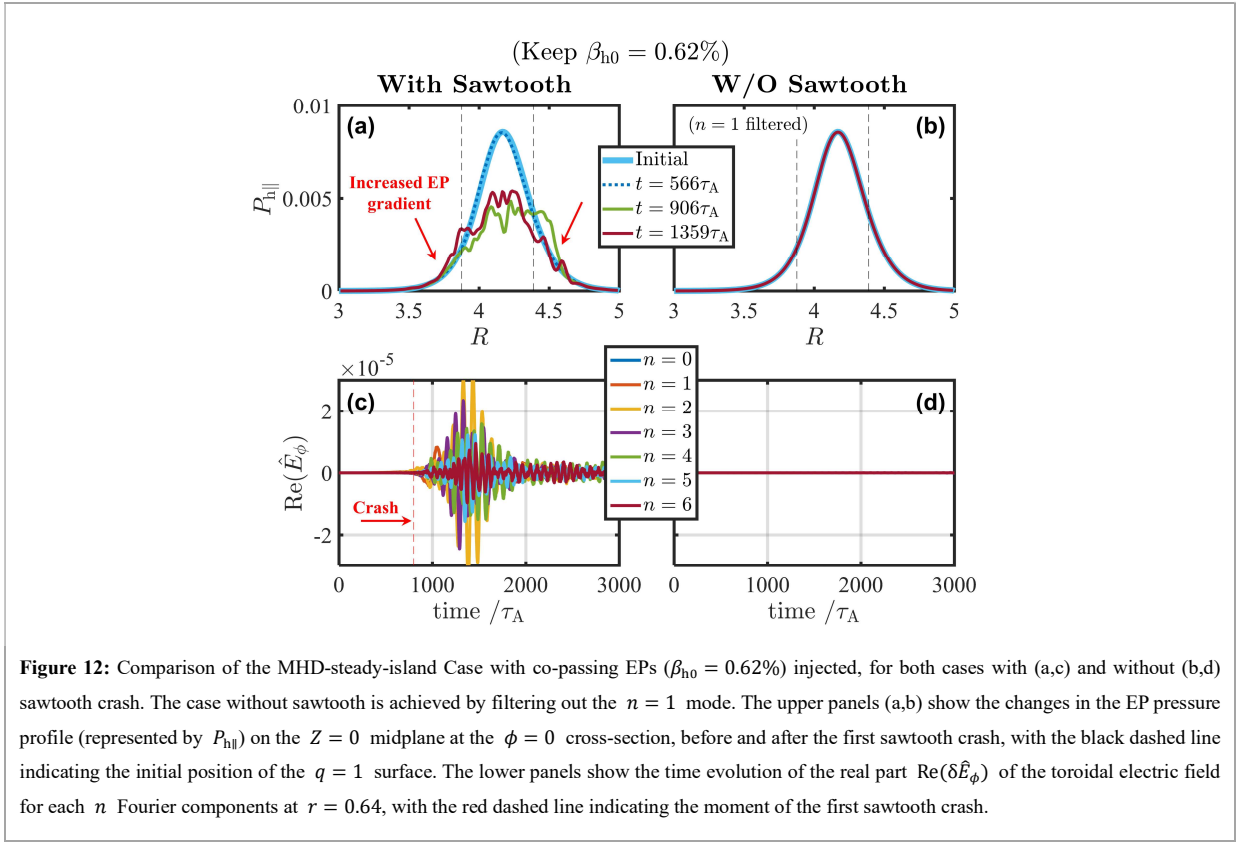

**Figure 12:** Comparison of the MHD-steady-island Case with co-passing EPs ($\beta_{h0} = 0.62\%$) injected, for both cases with (a,c) and without (b,d) sawtooth crash. The case without sawtooth is achieved by filtering out the $n = 1$ mode. The upper panels (a,b) show the changes in the EP pressure profile (represented by $P_{h\parallel}$) on the $Z = 0$ midplane at the $\phi = 0$ cross-section, before and after the first sawtooth crash, with the black dashed line indicating the initial position of the $q = 1$ surface. The lower panels show the time evolution of the real part $\mathrm{Re}(\delta\hat{E}_\phi)$ of the toroidal electric field for each $n$ Fourier components at $r = 0.64$, with the red dashed line indicating the moment of the first sawtooth crash.

Studies have shown that the burst of multi-mode TAEs and mode-mode interactions can lead to the so-called "TAE avalanche" events, which bring a significant increase in instability and rapid loss of EPs [96,97]. As a low-$n$, low-frequency mode, sawtooth crash indirectly triggers multi-mode TAEs through EP redistribution, suggesting that sawtooth oscillations may play an important role in triggering the "avalanche" of EPs.

### 6.2 Resonant Tearing Mode Excitation in the Presence of Steady-State 1/1 Island

In Sections 5.1 and 5.4, we observed that injecting counter-passing EPs tends to suppress sawtooth oscillations, leading the system toward a steady-state. However, in the later stages of the evolution, the total kinetic energy increases (see Figures 5(d,e) and 9(d,g)), indicating the excitation of new modes in the presence of counter-passing EPs and a steady-state $1/1$ magnetic island.

Poincaré plots of magnetic field lines reveal that this kinetic energy growth arises from the sudden expansion of the $m/n = 2/1$ magnetic island in the later stages of system evolution, i.e., the excitation of the $2/1$ tearing mode, as shown in Figure 13. We refer to this mode as the "resonant tearing mode" (r-TM), because, in the selected initial equilibrium, the $2/1$ rational surface has $\Delta'_{2/1} < 0$, and thus the $2/1$ tearing mode is theoretically MHD-stable; however, the large $2/1$ magnetic island is excited due to the resonance effect with the steady-state $1/1$ magnetic island in the presence of counter-passing EPs. Further investigation of the physical mechanisms of the r-TM will be presented in a separate study.

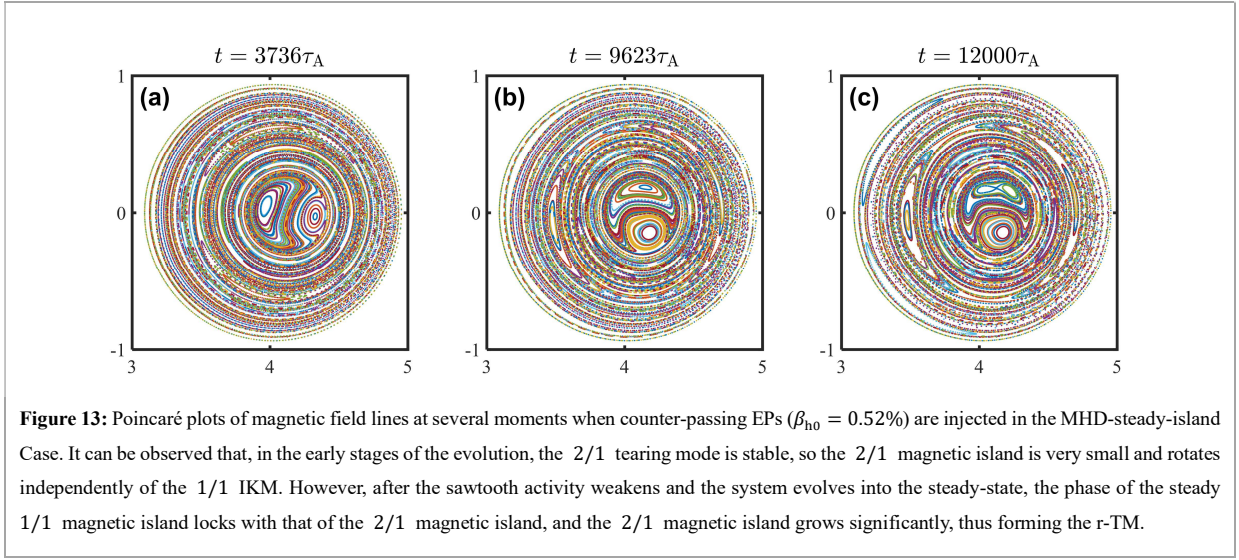

**Figure 13:** Poincaré plots of magnetic field lines at several moments when counter-passing EPs ($\beta_{h0} = 0.52\%$) are injected in the MHD-steady-island Case. It can be observed that, in the early stages of the evolution, the $2/1$ tearing mode is stable, so the $2/1$ magnetic island is very small and rotates independently of the $1/1$ IKM. However, after the sawtooth activity weakens and the system evolves into the steady-state, the phase of the steady $1/1$ magnetic island locks with that of the $2/1$ magnetic island, and the $2/1$ magnetic island grows significantly, thus forming the r-TM.

Recently, r-TMs associated with sawtooth crashes have also been observed on HL-2A, and simulations using the M3D-K code confirmed that they are also excited by counter-passing EPs [98]. Although their mechanism is not the same as the r-TM we observed, these findings collectively suggests that while counter-passing EPs can help induce small sawteeth to avoid NTMs, they may also lead to the potential excitation of r-TMs and the formation of large $2/1$ islands. This indicates that counter-passing EPs may also pose a confinement risk, which should be considered in future fusion experiments.

### 6.3 Influences of EP Energy and Safety Factor on Redistribution

In previous studies, we considered the standard case with fixed EP energy (birth speed $v_0 = 0.7v_A$). However, EP transport and redistribution weaken as EP energy increases, because faster EPs tend to induce more pronounced toroidal precession and larger orbit width, thereby decoupling from the motion of the magnetic surface [39-44]. In Figure 14, we examine the evolution of the kinetic energy and the EP redistribution for different EP energies ($v_0 = 0.35v_A$, $v_0 = 0.7v_A$, and $v_0 = 1.4v_A$). It is evident that the EP redistribution decreases with increasing EP energy. For the same energy, co-passing EPs are more easily redistributed than counter-passing EPs, as explained in Reference [44]. Notably, the degree of the EP redistribution is directly determined by the shift $(\Delta r)_\mathrm{d} \approx \frac{v_\mathrm{d} B}{v_\parallel B_\theta} r \approx q(r)\varrho_\mathrm{h}$ (where $v_\mathrm{d}$ is the magnetic drift speed and $B_\theta$ is the poloidal magnetic field) of the EP trajectory in the toroidal magnetic field, which is proportional to the EP Larmor radius $\varrho_\mathrm{h} = m_\mathrm{h} v / q_\mathrm{h} B$. This means that if the EP charge-to-mass ratio $(q_\mathrm{h}/m_\mathrm{h})$ increased proportionally with speed, the impact of the EP energy on the redistribution will be offset. The effects of co-passing and counter-passing EPs on sawtooth behavior are generally consistent with previous conclusions, but the r-TMs excited by injecting lower-energy counter-passing EPs $(v_0 = 0.35 v_\mathrm{A})$ are less pronounced.

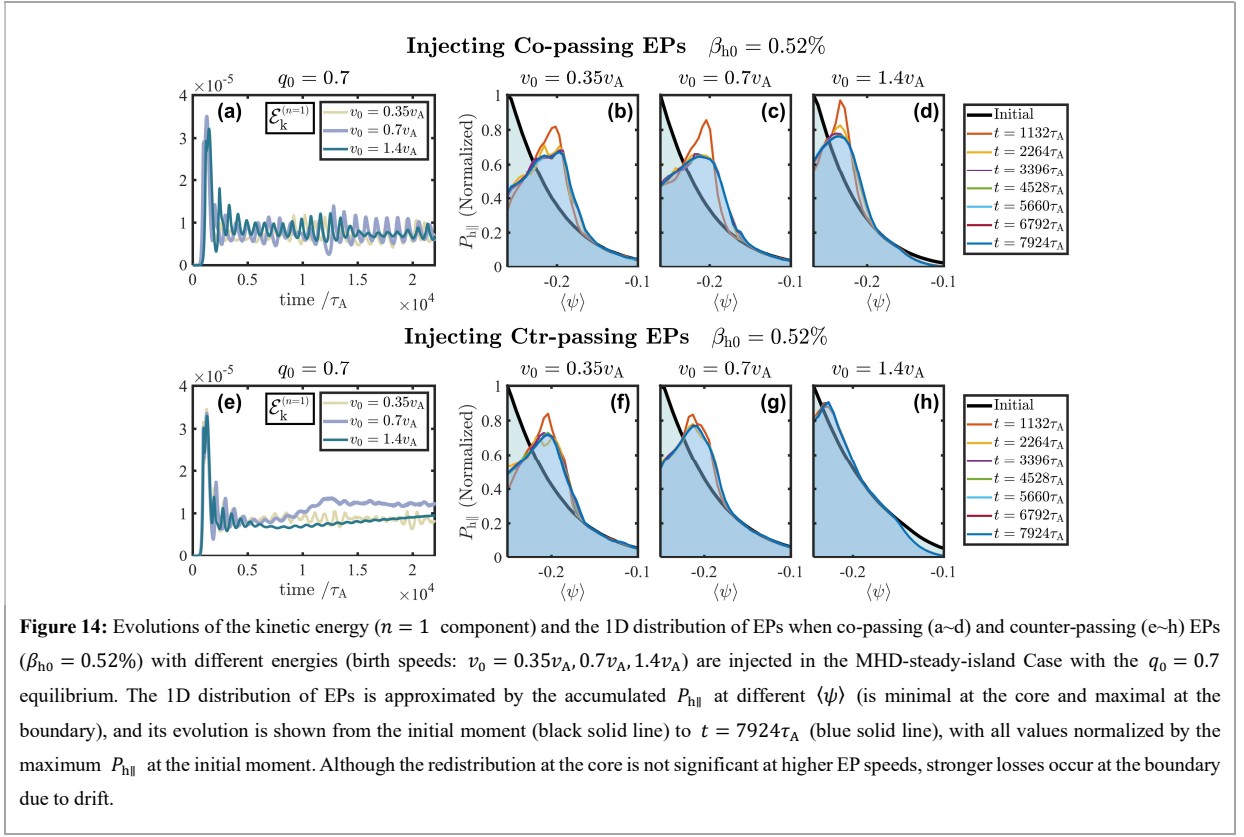

**Figure 14:** Evolutions of the kinetic energy ($n = 1$ component) and the 1D distribution of EPs when co-passing (a~d) and counter-passing (e~h) EPs ($\beta_{\mathrm{h}0} = 0.52\%$) with different energies (birth speeds: $v_0 = 0.35 v_\mathrm{A}, 0.7 v_\mathrm{A}, 1.4 v_\mathrm{A}$) are injected in the MHD-steady-island Case with the $q_0 = 0.7$ equilibrium. The 1D distribution of EPs is approximated by the accumulated $P_\mathrm{hl}$ at different $\langle \psi \rangle$ (is minimal at the core and maximal at the boundary), and its evolution is shown from the initial moment (black solid line) to $t = 7924\tau_\mathrm{A}$ (blue solid line), with all values normalized by the maximum $P_\mathrm{hl}$ at the initial moment. Although the redistribution at the core is not significant at higher EP speeds, stronger losses occur at the boundary due to drift.

The influence of the safety factor profile is also noteworthy. To compare with the previously used equilibrium with $q_0 = 0.7$, we consider equilibria with $q_0 = 0.8$ and $q_0 = 0.9$ (with all other parameters unchanged except for the core $q$ profile, as shown in Figure 15(a)). Figure 15(b~e) show the Poincaré sections of magnetic field lines before and after the first crash for the $q_0 = 0.9$ equilibrium with injected passing EPs. It is evident that for smaller $|1 - q_0|$, there are fewer regions of stochastic magnetic fields at the edge of the 1/1 island and outside the $q = 1$ surface (appearing only in panel (c)), contrasting sharply with the large stochastic field areas shown in Figure 6. The effect of larger $|1 - q_0|$ on the stochastic behavior of magnetic field has been confirmed, which is related to the excitation of low-order rational surface resonances [83,99,100]. The presence of stochastic fields significantly enhances EP transport [44,101,102], and thus, the larger $|1 - q_0|$ is, the more EPs redistribute, as shown in Figure 16. The change in $q_0$ does not alter the qualitative conclusions about sawtooth behaviors, but a higher $q_0$ results in a smaller sawtooth amplitude.

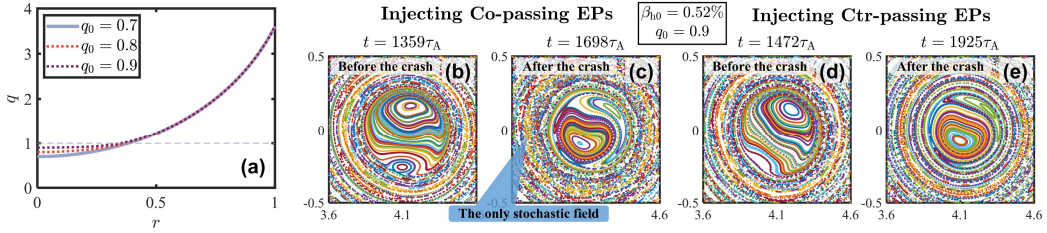

**Figure 15:** Initial equilibrium safety factor profiles after changing $q_0$ (a) and the evolution of the magnetic field structure before and after the first crash when co-passing (b,c) and counter-passing (d,e) EPs ($\beta_{h0} = 0.52\%$) are injected in the MHD-steady-island Case with the $q_0 = 0.9$ equilibrium. Stochastic fields almost do not appear, except observed outside the $q = 1$ rational surface in panel (c).

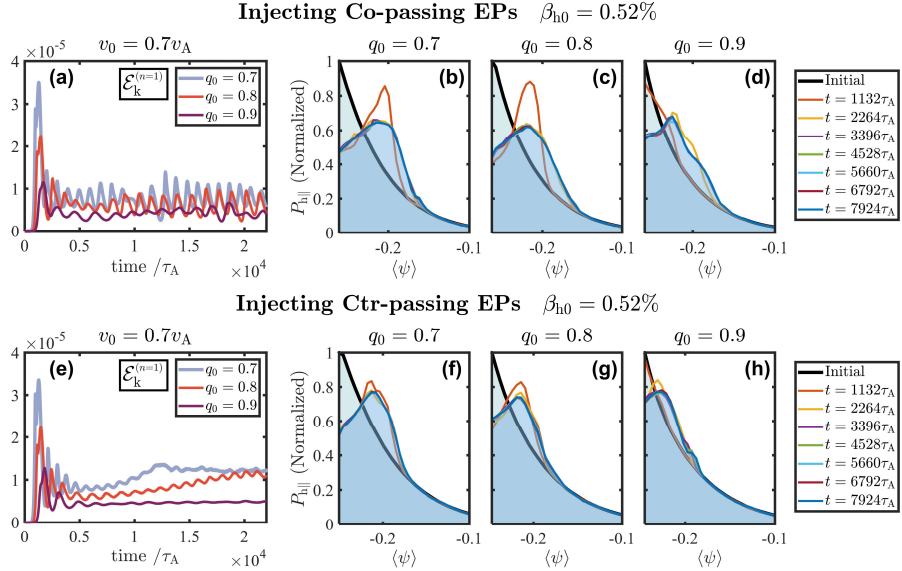

**Figure 16:** Evolutions of the kinetic energy ($n = 1$ component) and the 1D distribution of EPs when co-passing (a~d) and counter-passing (e~h) EPs ($\beta_{h0} = 0.52\%$) with $v_0 = 0.7v_A$ are injected in the MHD-steady-island Case with different initial equilibrium ($q_0 = 0.7, 0.8, 0.9$). The 1D distribution of EPs (b~d, f~h) is approximated by the accumulated $P_{hil}$ at different $\langle\psi\rangle$ (is minimal at the core and maximal at the boundary), and its evolution is shown from the initial moment (black solid line) to $t = 7924\tau_A$ (blue solid line), with all values normalized by the maximum $P_{hil}$ at the initial moment. The injection of co-passing EPs is more likely to generate a stochastic field outside the $q = 1$ surface compared to the injection of counter-passing EPs, so co-passing EPs can be transported to more outward locations.

## 6.4 Necessity of Multi-$n$ Simulations

Finally, we briefly emphasize the necessity of multi-$n$ simulations. Since sawtooth behavior is inherently nonlinear [35], its dynamics are strongly influenced by mode-mode coupling, and linear simulations or single-$n$ simulations cannot fully capture sawtooth dynamics. For MHD contributions, retaining as many $n$ components as possible is clearly essential [82,102,103]. If only a single-$n$ MHD evolution is considered, sawtooth crashes may not even occur. In this study, we have retained all modes from $n = 0$ to $n = 7$ for the MHD contribution.

Particular attention must be given to the treatment of EP contributions. Many previous studies on EP physics have coupled only the single-$n$ components (e.g., $n = 1$) of the EP contribution, which is suitable for studying isolated instabilities but not complex nonlinear processes like sawtooth oscillations. Specifically, EP redistribution caused by sawtooth crashes leads to a dominant $n = 0$ component in the perturbation of the EP distribution function during the nonlinear stage, significantly affecting the results. In our simulations, we also retained all $n$ modes from $n = 0$ to $n = 7$ for the EP contribution. Figure 17 shows the impact of retaining different $n$ components of the EP contribution on the energy curve. While the overall conclusions remain unchanged (i.e., co-passing EPs favor sawtooth re-excitation, counter-passing EPs promote steady-state formation and excite r-TMs), selectively filtering out components leads to notable differences in simulation results, (e.g., the sawtooth period in panel (a) and the early growth of the $n = 0$ zonal flow in panel (e)). Moreover, filtering high-$n$ components prevents the observation of multi-$n$ TAE excitation due to sawtooth crashes. This comparison highlights the necessity of multi-$n$ nonlinear simulations using initial-value codes for sawtooth research.

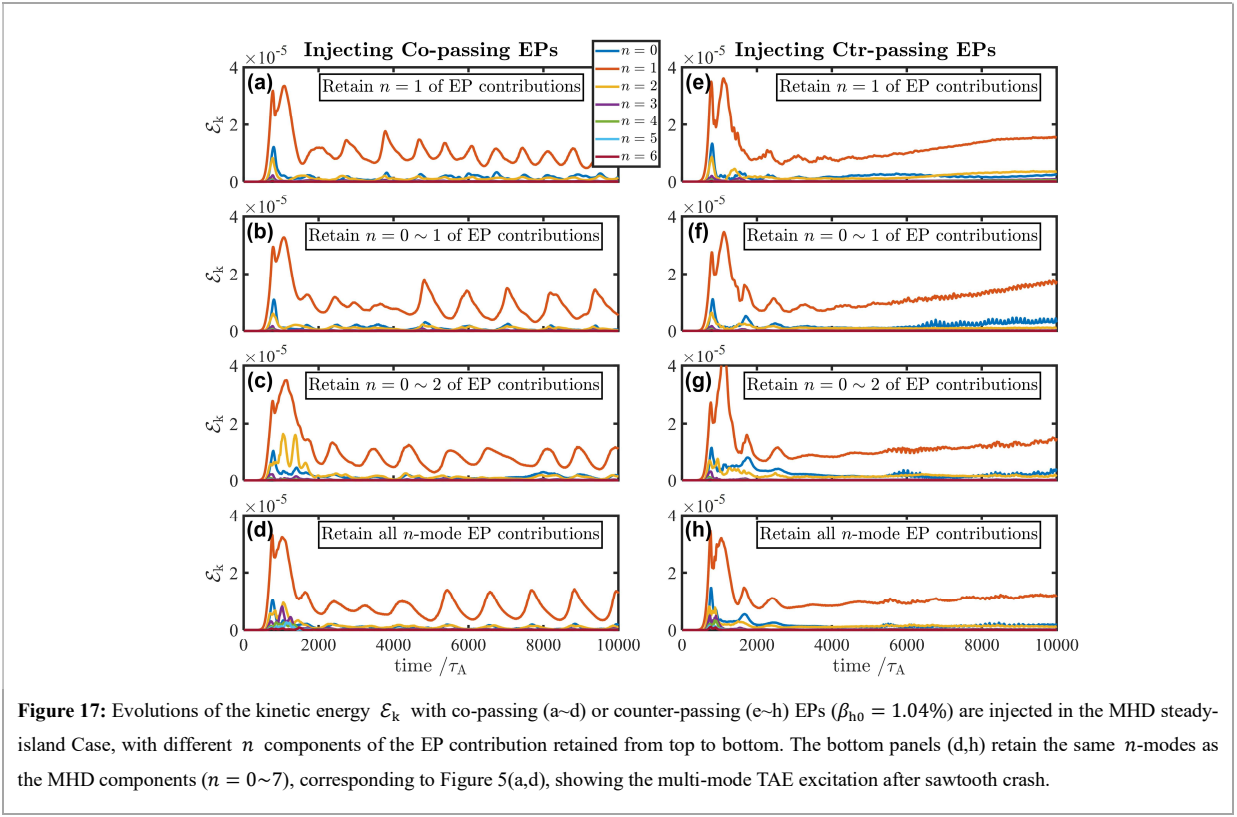

**Figure 17:** Evolutions of the kinetic energy $\mathcal{E}_k$ with co-passing (a~d) or counter-passing (e~h) EPs ($\beta_{h0} = 1.04\%$) are injected in the MHD steady-island Case, with different $n$ components of the EP contribution retained from top to bottom. The bottom panels (d,h) retain the same $n$-modes as the MHD components ($n = 0\sim7$), corresponding to Figure 5(a,d), showing the multi-mode TAE excitation after sawtooth crash.

## 7.    Summary and Discussion

Sawtooth oscillations in Tokamaks exhibit significant interactions with EPs. Sawtooth crashes can lead to EP transport and redistribution, which, in turn, may affect sawtooth behavior by stabilizing or destabilizing the IKM. However, long-term nonlinear simulations using initial-value codes on this topic remain scarce. In this study, we use the MHD-kinetic hybrid code CLT-K to provide a self-consistent analysis of this interaction. We find that EPs can influence the radial residual flow after each sawtooth cycle by altering the sawtooth period, and that radial residual flow directly affects the reconnection rate at the X-point, thereby influencing the sawtooth type. Shorter sawtooth periods correspond to stronger residual flows, which favor the system reaching a steady-state, while longer periods promote the sawtooth re-excitation. Our conclusions are qualitatively consistent with both theoretical predictions and experimental observations, and the main results are summarized in Table 2.

Additionally, we confirm that for long-term nonlinear simulations, multi-mode simulations are necessary as they reveal richer physics. For instance, EP redistribution caused by sawtooth crashes can lower the excitation threshold of global high-$n$ TAEs. Furthermore, although counter-passing EPs are typically effective in controlling the sawtooth period and preventing NTMs by inducing small sawteeth, once a steady-state $1/1$ magnetic island is formed, they may instead drive the excitation of the $2/1$ r-TM and form the large magnetic island. This dual effect warrants attention in future fusion experiments. The influence of EP energy and safety factor profiles on redistribution has also been studied, with particular focus on the interesting role played by stochastic fields.

**Table 2:** Summary of simulation results on the impact of passing EPs on sawtooth oscillation type transitions.

| On-axis Injection | MHD-normal-sawtooth Case | MHD-small-sawtooth Case | MHD-steady-island Case |
|---|---|---|---|
| **Co-passing EPs** | Maintain normal sawtooth **(Bad)** | Extend period **(Bad)** (may evolve into normal sawtooth) | Re-excite new small sawtooth **(Good)** |
| **Ctr-passing EPs** | Can evolve into small sawtooth **(Good)** | Shorten period **(Good)** (may evolve into steady-state) | Maintain steady-state but excite r-TMs **(Bad)** |

The main limitation of this work lies in the use of a simplified single-fluid model for the bulk plasma, excluding more complex effects such as two-fluid effects, which could significantly influence sawtooth behavior. Halpern et al. [104] showed that accounting for diamagnetic drift effects can reduce sawtooth crash time, aligning results more closely with experimental observations. The latest version of the CLT-K code now supports modules for diamagnetic drift [105], Hall effect [106], shear flows [107], and bootstrap currents [108], and their impact will be explored in future work. Additionally, we plan to incorporate more self-consistent heat and current sources, along with source and sink terms for EPs. This study focuses on passing EPs (the primary component in tangential NBI injection), with future research expanding to trapped and isotropic-distributed EPs (e.g., fusion alpha particles).

This paper focuses on the interaction between EP redistribution and sawtooth oscillations at low EP fractions. It is well known that when the EP fraction reaches a certain threshold, high-frequency instabilities such as fishbones [109] and Alfvén eigenmodes [110] can be triggered. A series of theoretical studies and experimental observations [31,101,111-113] have suggested that these high-frequency instabilities may coexist with or influence sawteeth (e.g., sawtooth-fishbone interactions). However, long-term nonlinear self-consistent simulations using initial-value codes on this topic are still limited. We sincerely look forward to potential collaborations and further insights into this subject, as understanding the underlying physical mechanisms and exploring the parameter space will greatly benefit the operation of future fusion devices such as ITER.

## Acknowledgements

We would like to express our sincere gratitude to the representatives at the TMEP 2025 conference for their constructive discussions and valuable suggestions, and we also acknowledge the Max Planck Institute for Plasma Physics for supporting the presentation of this work at the TMEP 2025 conference. This work is supported by the National MCF Energy R&D Program No. 2022YFE03100001 and 2019YFE03030004.